\documentclass[%
reprint,
superscriptaddress,
bibnotes,
amsmath,amssymb,
aps,
prl,
longbibliography,
]{revtex4-2}

\usepackage{graphicx}
\usepackage{dcolumn}
\usepackage{bm}
\usepackage{chemformula}
\usepackage{hyperref}

\begin{document}


\title{Easy-plane multi-$\mathbf{q}$ magnetic ground state of \ch{Na_3Co_2SbO_6}}

\author{Yuchen~Gu}
\affiliation{International Center for Quantum Materials, School of Physics, Peking University, Beijing 100871, China}
\author{Xintong~Li}
\affiliation{International Center for Quantum Materials, School of Physics, Peking University, Beijing 100871, China}
\affiliation{Beijing National Laboratory for Condensed Matter Physics, and Institute of Physics, Chinese Academy of Sciences, Beijing 100190, China}
\author{Yue~Chen}
\affiliation{International Center for Quantum Materials, School of Physics, Peking University, Beijing 100871, China}
\author{Kazuki~Iida}
\affiliation{Neutron Science and Technology Center, Comprehensive Research Organization for Science and Society, Tokai, Ibaraki 319-1106, Japan}
\author{Akiko~Nakao}
\affiliation{Neutron Science and Technology Center, Comprehensive Research Organization for Science and Society, Tokai, Ibaraki 319-1106, Japan}
\author{Koji~Munakata}
\affiliation{Neutron Science and Technology Center, Comprehensive Research Organization for Science and Society, Tokai, Ibaraki 319-1106, Japan}
\author{V.~Ovidiu~Garlea}
\affiliation{Neutron Scattering Division, Oak Ridge National Laboratory, Oak Ridge, Tennessee 37831, USA}
\author{Yangmu~Li}
\affiliation{Beijing National Laboratory for Condensed Matter Physics, and Institute of Physics, Chinese Academy of Sciences, Beijing 100190, China}
\affiliation{Condensed Matter Physics and Materials Science Division, Brookhaven National Laboratory, Upton, New York 11973, USA}
\affiliation{School of Physical Sciences, University of Chinese Academy of Sciences, Beijing 100049, China}
\author{Guochu~Deng}
\affiliation{Australian Centre for Neutron Scattering, Australian Nuclear Science and Technology Organisation, Lucas Heights NSW-2234, Australia}
\author{I.~A.~Zaliznyak}
\email{zaliznyak@bnl.gov}
\affiliation{Condensed Matter Physics and Materials Science Division, Brookhaven National Laboratory, Upton, New York 11973, USA}
\author{J.~M.~Tranquada}
\affiliation{Condensed Matter Physics and Materials Science Division, Brookhaven National Laboratory, Upton, New York 11973, USA}
\author{Yuan~Li}
\email{yuan.li@pku.edu.cn}
\affiliation{International Center for Quantum Materials, School of Physics, Peking University, Beijing 100871, China}
\date{\today}
\begin{abstract}

\ch{Na_3Co_2SbO_6} is a potential Kitaev magnet with a monoclinic layered crystal structure. Recent investigations of the $C_3$-symmetric sister compound \ch{Na_2Co_2TeO_6} have uncovered a unique triple-$\mathbf{q}$ magnetic ground state, as opposed to a single-$\mathbf{q}$ (zigzag) one, prompting us to examine the influence of the reduced structural symmetry of \ch{Na_3Co_2SbO_6} on its ground state. Neutron diffraction data obtained on a twin-free crystal reveal that the ground state remains a multi-$\mathbf{q}$ state, despite the system's strong in-plane anisotropy. This robustness of multi-$\mathbf{q}$ orders suggests that they are driven by a common mechanism in the honeycomb cobaltates, such as higher-order magnetic interactions. Spin-polarized neutron diffraction results show that the ordered moments are entirely in-plane, with each staggered component orthogonal to the propagating wave vector. The inferred ground state favors a so-called XXZ easy-plane anisotropic starting point for the microscopic model over a Kitaev one, and features unequal ordered moments reduced by strong quantum fluctuations.

\end{abstract}

\maketitle

Magnetic frustration arises from competing interactions between localized magnetic moments, or spins, leading to a vast degeneracy of classical ground states and suppressed order formation in quantum systems \cite{BalentsNature2010, ZhouRMP2017, BroholmScience2020}. Acquiring precise knowledge of the order parameter can provide valuable insights when a frustrated magnet attains order. However, obtaining such information can be challenging. The Kitaev honeycomb model \cite{KitaevAP2006} has garnered interest due to its unique magnetic frustration properties, exact quantum solvability, and potential applications in topological quantum computation \cite{TakagiNRP2019}. Significant research progress in materializing the Kitaev model has been made \cite{JackeliPRL2009, ChaloupkaPRL2010, PlumbPRB2014, TakagiNRP2019, MotomeJPCM2020, TrebstPRRSP2022}, with $3d$ cobaltates recently emerging as promising materials \cite{LiuPRB2018, SanoPRB2018, LiuPRL2020, Kim_2021, LiuIJMPB2021}.

Two key factors driving research interest in honeycomb cobaltates are the appealing theoretical expectation of weak non-Kitaev interactions \cite{LiuPRB2018, SanoPRB2018} and the growth of large, high-quality single crystals \cite{XiaoCGD2019, YanPRM2019, YaoPRB2020, ZhongSA2020}. However, some cobaltates have recently been argued to be better described as easy-plane anisotropic (XXZ) rather than Kitaev magnets \cite{HalloranPNAS2023, WinterJOP2022, DasPRB2021, LiuPRB2023}. The compound \ch{Na_3Co_2SbO_6} (NCSO) has nevertheless been considered to exhibit significant Kitaev interactions \cite{LiuPRB2023} and potential spin liquid behavior \cite{LiuPRL2020,PandeyPRB2022}. Furthermore, NCSO is a structurally well-defined and clean material \cite{LiPRX2022}, which are traits making it valuable for in-depth studies aiming to avoid the structural complications recently found in $\alpha$-RuCl$_3$ \cite{Zhang2023Arxiv} and \ch{Na_2Co_2TeO_6} \cite{DufaultArxiv2023}.

\begin{figure}[!b]
\includegraphics[width=2.6in]{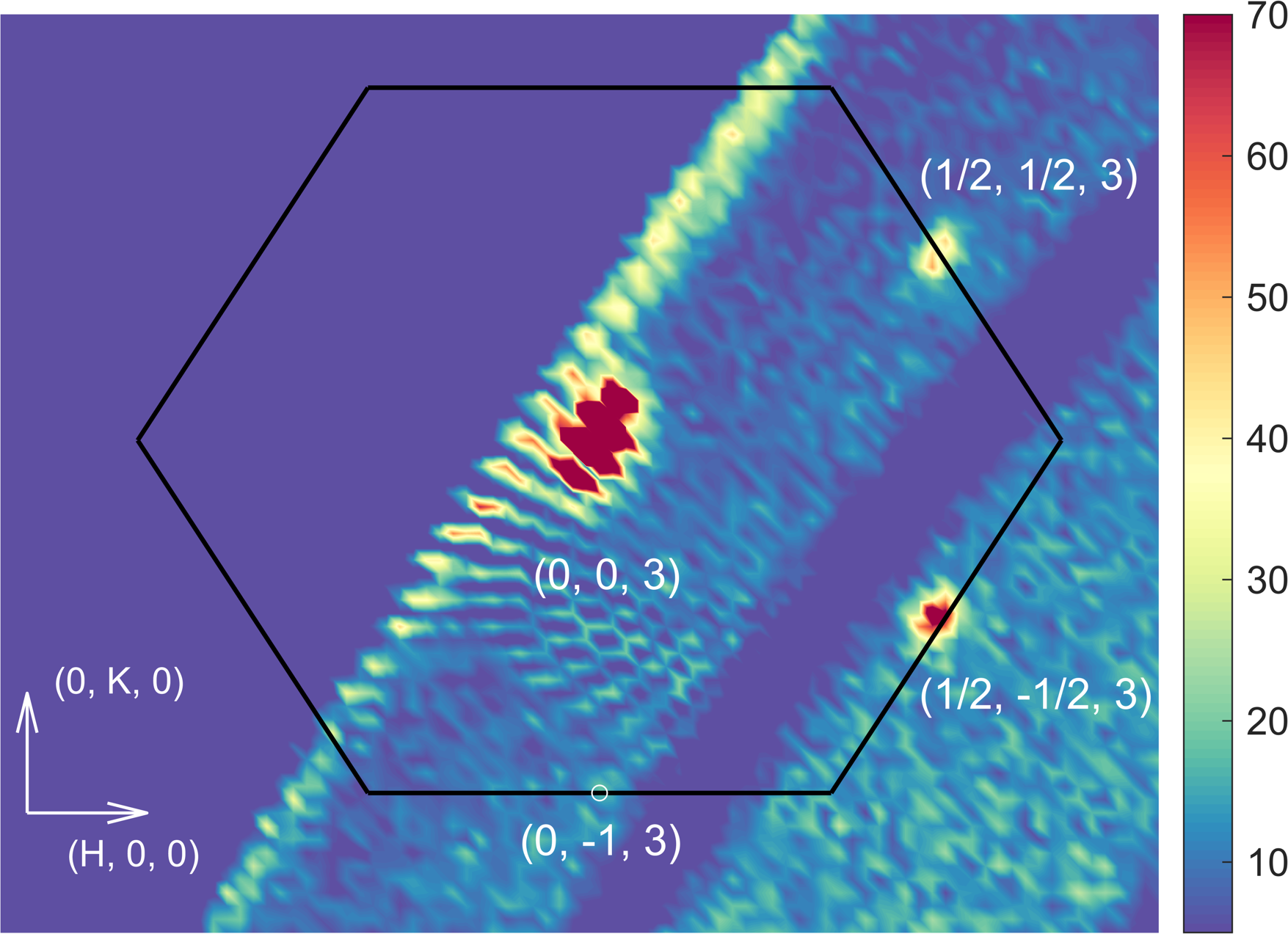}
\caption{Neutron diffraction on a twin-free crystal measured on SENJU \cite{OhharaJAC2016} at 2 K and in zero field. Data are averaged from $(H, K, 2.9)$ to $(H, K, 3.1)$ in reciprocal lattice units (r.l.u.). Hexagon indicates the 2D Brillouin zone (BZ), which is elongated along $(1,0,0)$ due to the monoclinic inclination of the $c$ axis ($\beta=108.6^\circ$). Magnetic peaks are observed at $(H,K)=(1/2, \pm1/2)$ but not at $(0,-1)$. Washboard-like texture is due to small gaps between detectors.
}
\label{fig1}
\end{figure}

\begin{figure*}
\includegraphics[width=\textwidth]{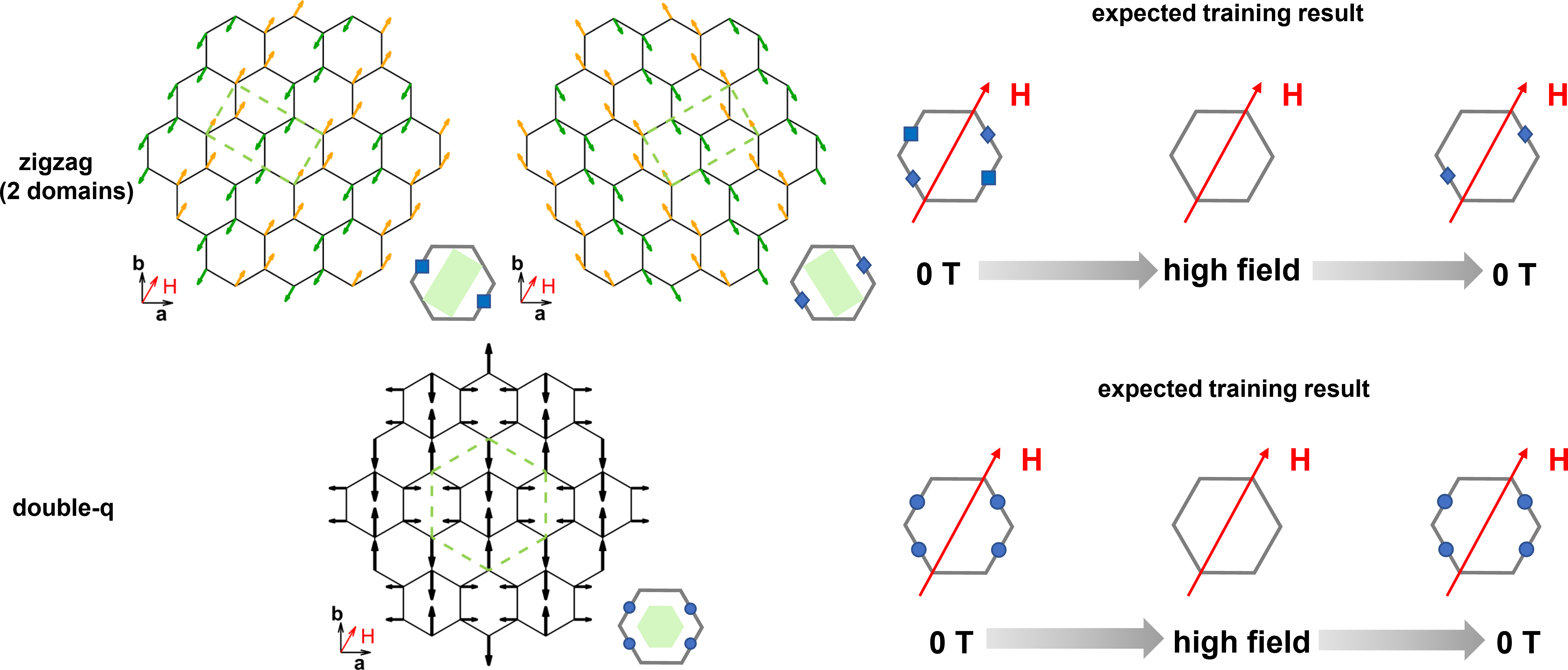}
\caption{Left half: Schematic spin patterns of two types of zigzag domains and of the multi-$\mathbf{q}$ order. The spin orientations are constrained by our spin-polarized diffraction data in Fig.~\ref{fig4}. Dashed lines indicate a magnetic primitive cell. Lower-left insets show the applied-field ($\mathbf{H}$) direction in the crystallographic coordinate system, and lower-right the 2D structural (grey hexagon) and magnetic (filled polygon) Brillouin zones and locations of the magnetic Bragg peaks. Right half: Expected field-training results observable by magnetic neutron diffraction, under the zigzag and multi-$\mathbf{q}$ scenarios. See text for detailed explanation.}
\label{fig2}
\end{figure*}

In candidate Kitaev materials, including cobaltates, the antiferromagnetic order of zigzag ferromagnetic chains, dubbed the zigzag order, is widely regarded as the predominant form of magnetic ground state \cite{CaoPRB2016, LiuPRB2011, LefrancoisPRB2016, BeraPRB2017, YanPRM2019}. The zigzag order is characterized by a single propagating wave vector ($\mathbf{q}$) at one of the $M$-points of the hexagonal Brillouin zone (BZ). However, recent research on \ch{Na_2Co_2TeO_6} has unveiled a surprising triple-$\mathbf{q}$ ordered state \cite{ChenPRB2021, KrugerArxiv2022, YaoPRL2022, YaoArxiv2022, LeePRB2021, KikuchiArxiv2022}. Despite ongoing debate about its relevance \cite{LinNC2021, SongvilayPRB2020, KimJOP2021}, the triple-$\mathbf{q}$ state can be identified as a superposition of three single-$\mathbf{q}$ zigzag components rotated by 120 degrees from one another \cite{ChenPRB2021,JanssenPRL2016}. Recent theoretical studies suggest that a multi-$\mathbf{q}$ state can become energetically favorable over a zigzag state when higher-order spin interactions are considered \cite{KrugerArxiv2022,PohlePRB2023,WangarXiv2023}. While the multi-$\mathbf{q}$ state in \ch{Na_2Co_2TeO_6} preserves the lattice $C_3$ rotational symmetry about the $c$-axis, it remains unclear whether the state is necessarily $C_3$-symmetric or can be stable even in a lower-symmetry setting, potentially due to the prominence of higher-order spin interactions.

In this Letter, our investigation of NCSO addresses two crucial questions: whether the system is better characterized as an XXZ rather than a Kitaev magnet, and whether the lack of $C_3$ symmetry is compatible with the formation of multi-$\mathbf{q}$ order. Using neutron diffraction on a twin-free crystal, we reveal the presence of two, rather than one, or three, zigzag-like antiferromagnetic components in zero field. We show that the two components belong to the same multi-$\mathbf{q}$ (double-$\mathbf{q}$) order parameter, the critical evidence being that their signal ratio remains unchanged after the system is trained by strong in-plane magnetic fields along a low-symmetry direction. Spin-polarized neutron diffraction further demonstrates that the staggered spins in each zigzag component lie entirely in-plane and perpendicular to the staggered wave vector, which is more compatible with the XXZ model than the Kitaev model. Superimposing the components as revealed by the spin-polarized diffraction data yields a 2D non-collinear spin pattern with unequal moment sizes. Since the reduction of classically ordered moments is a hallmark of quantum fluctuations, our results render NCSO as a promising system for exploring spin-liquid physics.

The space-group symmetry of NCSO is $C2/m$ (No. 12) \cite{YanPRM2019,LiPRX2022}, the same as that of the high-temperature structure of $\alpha$-\ch{RuCl_3} \cite{JohnsonPRB2015,CaoPRB2016,Zhang2023Arxiv}. Figure~\ref{fig1} displays our neutron diffraction data obtained on a white-beam diffraction instrument \cite{OhharaJAC2016} from a 6 mg twin-free crystal \cite{LiPRX2022}, covering the $(H,K,3)$ reciprocal plane. In zero field, we observe magnetic Bragg peaks at only two of the three $M$-points of the pseudo-hexagonal BZ: at $(1/2,1/2,3)$ and $(1/2,-1/2,3)$, but not at $(0,-1,3)$. This finding is consistent with previous reports using twinned crystals \cite{YanPRM2019,LiPRX2022}. Our complete dataset verifies the absence of magnetic peaks at $(0,\pm1,L)$ or other symmetry-related positions in higher-index 2D BZs over a wide range of $L$ values.

\begin{figure*}[!ht]
\includegraphics[width=\textwidth]{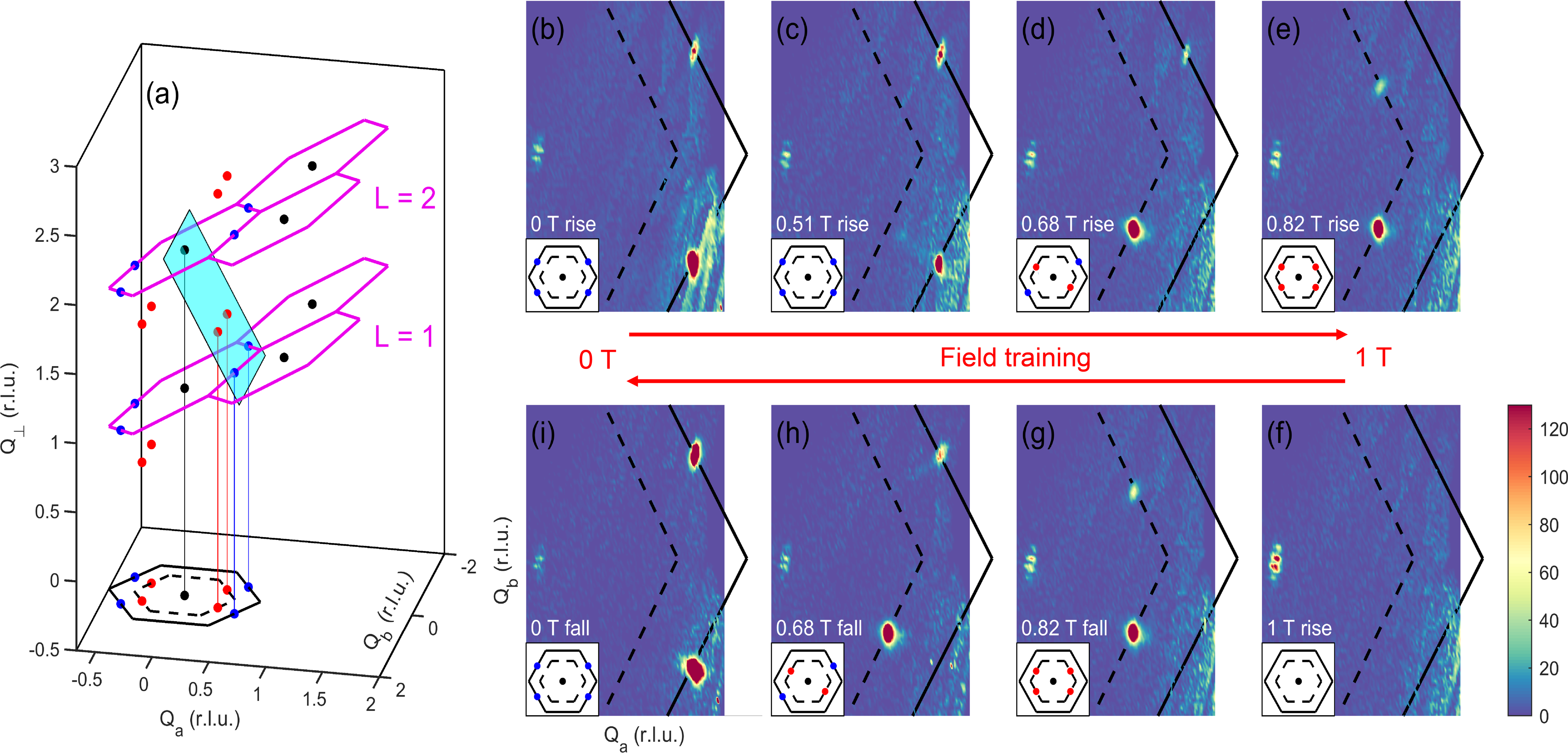}
\caption{(a) Schematic of diffraction peaks (filled spheres) in reciprocal space. An orthogonal coordinate system is chosen to include the $ab$ plane and its normal direction, such that magnetic diffraction peaks are seen to project onto the $M$-points (blue) and the ``$\frac{2}{3}M$-points'' (red) of the 2D BZ \cite{LiPRX2022}. Magenta hexagons indicate 3D monoclinic BZ boundaries at $L=1$ and 2. Cyan plane indicates the momentum slice displayed in (b-i), where the data are obtained on SENJU \cite{OhharaJAC2016} with the field applied and removed in the displayed order. The field direction is in the $ab$ plane and at 30 degrees from $b$, same as in Fig.~\ref{fig2}. The coordinate system next to (i) indicates the cyan plane's coordinates projected into the $Q_a$-$Q_b$ plane at the bottom of (a). Colored spheres in the inset of (b-d) indicate 2D magnetic peak positions. The observed magnetic peaks, upon their first appearance, have the following $(H,K,L)$ indices: (b-c) $(1/2,1/2,1)$ and $(1/2,-1/2,1)$; (d) $(1/3,-1/3,4/3)$; (e) $(1/3,1/3,4/3)$. }
\label{fig3}
\end{figure*}

Figure~\ref{fig2} demonstrates that the above result is in principle consistent with both a zigzag and a multi-$\mathbf{q}$ ordered state. In the zigzag scenario, magnetic Bragg peaks at $(H,K)=\pm(1/2,1/2)$ and $\pm(1/2,-1/2)$ originate from two types of domains (excluding time-reversal). They are related by a 180-degree rotation about the $b$ axis ($C_{2,b}$), which is a crystallographic symmetry, and are thus expected to coexist in a macroscopic sample. In the multi-$\mathbf{q}$ scenario, the ordering pattern can be regarded as a superposition of the two zigzag patterns just considered, with all magnetic Bragg peaks emerging simultaneously. The non-zero structure factors at only two instead of all three $M$-points are consistent with the system's monoclinic symmetry, where the two $M$-points form a symmetry-enforced wave vector star. The lack of a diffraction peak at the third $M$-point marks the absence of higher harmonics of the magnetization modulations. It makes the spin pattern (Fig.~\ref{fig2}) deviate from the $C_3$-symmetric one proposed in \cite{ChenPRB2021}, likely owing to NCSO's strong in-plane anisotropy \cite{LiPRX2022}.

\begin{figure}[!ht]
\includegraphics[width=2.7in]{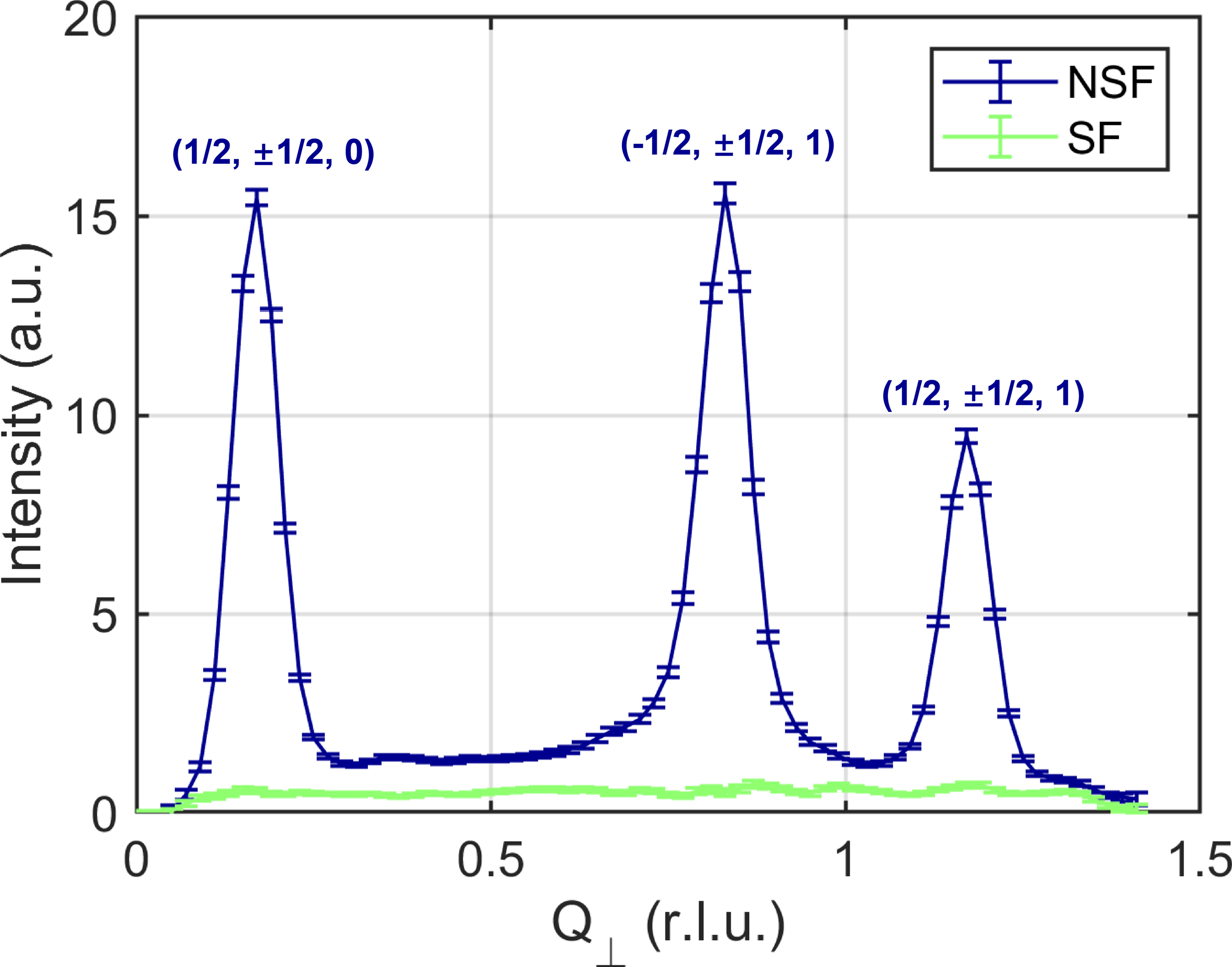}
\caption{Momentum scans perpendicular to the $ab$ plane, measured on HYSPEC \cite{ZaliznyakIOP2017} at 0.3 K with polarized neutrons on a twinned sample. Spin-flip (SF) and non-spin-flip (NSF) data have been corrected by the flipping ratio \cite{SM}. See Fig.~\ref{fig3}(a) for the definition of $Q_\perp$.
}
\label{fig4}
\end{figure}

To distinguish between the two-domain zigzag and the double-$\mathbf{q}$ scenarios, we study the impact of training the sample in an in-plane magnetic field applied in a direction rotated 30 degrees from the $b$ axis. Magnetic diffraction peaks are monitored in a momentum plane indicated by the cyan plane in Fig.~\ref{fig3}(a). Peaks will be identified by their in-plane components $(Q_a, Q_b)$. Before we discuss the data, it is useful to see why the magnetic field should affect the two types of zigzag domains differently. For the domains illustrated in Fig.~\ref{fig2}, the difference arises from the fact that one type of domain can slightly lower its energy in the field by spin canting towards the field direction, whereas the other type cannot. The locking between the spin and the wave vector directions, enforced by spin-orbit effects \cite{LiuPRB2018, SanoPRB2018, LiuPRL2020}, plays an important role here. Although we will later show that the specific spin orientations in Fig.~\ref{fig2} are corroborated by spin-polarized neutron diffraction data, we emphasize that the difference in the field's influence is generically enforced by (the lack of) symmetry: with the field applied, the $C_{2,b}$ symmetry connecting the two types of domains becomes broken, so there is no longer a symmetry to protect the domains' energy degeneracy. As an aside, while the zigzag domains proposed in Ref.~\onlinecite{YanPRM2019}, with all spins lying parallel to the $b$ axis regardless of the zigzag-chain orientation, might appear degenerate in the field, the degeneracy is at best coincidental and not symmetry-enforced (as shown below, the magnetic structure in Ref.~\onlinecite{YanPRM2019} is inconsistent with our data in Fig.~\ref{fig4}).

Figure~\ref{fig3} presents the result of our field-training experiment. We stress that the key observation here is not about unequal impacts on the two pairs of magnetic diffraction peaks when the field is on, but about the remnant effect of a sufficiently large field applied and then removed. With the locking between the spin and the wave vector directions, the zigzag scenario is expected to have one type of domain noticeably depopulated after training, whereas the multi-$\mathbf{q}$ scenario should definitely return to its original state. We have selected measurement field strengths matching the known phase boundaries at 0.53, 0.73, and 0.91 Tesla for our field direction \cite{LiPRX2022}. Fields above 0.91 T drive the system into a ferromagnetic state, eliminating antiferromagnetic diffraction peaks. The data in Fig.~\ref{fig3} confirm this understanding; in fact, all the previously identified phases and peak indexing from \cite{LiPRX2022} are corroborated in our experiment. For details, see the caption of Fig.~\ref{fig3}.

In a nutshell, our data reveal that the field-training has minimal impact. The crucial observation, comparing Fig.~\ref{fig3}(b) and (i), is that both of the $M$-point diffraction peaks remain present after training. It must be noted that the two magnetic peaks in Fig.~\ref{fig3}(b-e, g-i) consistently displayed unequal intensities due to a technical reason: the lower peak was in the horizontal scattering plane, while the upper peak was outside the scattering plane. Using a non-monochromated beam for diffraction, the differing scattering geometries led to neutrons of varying kinetic energies, or wavelengths $\lambda$, contributing to the peaks. The data shown in Fig.~\ref{fig3} are not corrected for the Lorentz-factor ($\propto d^2\lambda^2$ \cite{MichelsClarkJAC2016}, where the $d$-spacings of the two peaks are equal), which explains the stronger intensity of the lower peak in Fig.~\ref{fig3}(i). Additionally, the upper peak in Fig.~\ref{fig3}(b-d) [compared to (h-i)] appears particularly weak because it partially fell outside the detector boundary. This issue was identified and resolved (by slightly rotating the sample) in subsequent measurements. More information on our measurement conditions can be found in \cite{SM}.

We emphasize that the absence of significant training effects contrasts with the distinct influence of the field on the two zigzag components. The latter is evident from the multi-step switching of the associated diffraction peaks in Figs.~\ref{fig3}(c-e) and (g-i). As previously mentioned, this distinction is anticipated from a symmetry perspective, since the field breaks the $C_{2,b}$ symmetry connecting the two components. It therefore manifests the significance of spin-orbit effects in NCSO. The divergent behaviors of the components upon the field application, along with their common and full recovery upon the field removal, confirm that the components form a multi-$\mathbf{q}$ magnetic state together. Further diffraction and magnetization measurements on twin-free crystals support these findings (Figs.~\ref{figS1} and \ref{figS2} in \cite{SM}).

Having obtained evidence of the multi-$\mathbf{q}$ nature, our next goal is to determine the ordered spin configuration. We first note that the multi-$\mathbf{q}$ order must comprise zigzag components that are collinear before their superposition, as non-collinearity would require a ``stripe'' component admixture, leading to non-zero structure factors at 2D wave vectors $(\pm 1/2, \pm 3/2)$ and $(\pm 1, 0)$ \cite{ChoiPRL2012}. However, such signals are absent in experiments \cite{LiPRX2022}. Thus, we focus on identifying the staggered spin direction within a single zigzag component, optimally achieved using spin-polarized neutron diffraction.

In Fig.~\ref{fig4}, we show diffraction measurements on an array of twinned crystals with a vertically spin-polarized incoming beam. The crystals' shared $c^*$ axis lies in the horizontal scattering plane, making it the reciprocal plane $(K,K,L)$, $(K,-K,L)$, or $(0,K,L)$, each for about one third of the crystals in the array. In this geometry, a zigzag component generates a series of diffraction peaks at the same 2D $M$-point in the scattering plane, such as $(1/2,1/2,L)$, where $L$ is an integer. Peaks from the first two aforementioned crystallographic twins are accessible, whereas the third has no magnetic reflections (Fig.~\ref{fig1}) in the scattering plane. An illustration of the scattering geometry can be found in Fig.~\ref{figS3} \cite{SM}. Spin components in the scattering plane produce spin-flip diffraction signals, while those perpendicular to the plane produce non-spin-flip signals. Figure \ref{fig4} shows that all magnetic diffractions are non-spin-flip, indicating that the spins in the zigzag components associated with measured $M$-points lie perfectly vertical in the laboratory frame, which is a direction in the honeycomb plane and perpendicular to the 2D $M$-point wave vectors, irrespective of the crystallographic twin origin of the signal. Consequently, we arrive at the zigzag components' spin orientations depicted in Fig.~\ref{fig2}. The full ordered spin configuration is obtained by superimposing the two zigzag components.

The double-$\mathbf{q}$ magnetic structure in Fig.~\ref{fig2} is non-collinear because of a particular choice of the \textit{collinear} spin orientations in the two constituent zigzag components. Importantly, as spins in the two components on the same sites are either 60 or 120 degrees apart, their vector superposition results in two distinct spin magnitudes ($\sqrt{3}:1$, each occupying half of the sites) in the double-$\mathbf{q}$ structure. An alternative way to view the magnetic structure is to decompose it into four sublattices made of third-nearest-neighbor bonds, which have recently been suggested to possess significant antiferromagnetic interactions \cite{YaoPRL2022,MaksimovPRB2022}. As shown by the dashed hexagon in Fig.~\ref{fig2}, each of the sub-lattices forms collinear N\'{e}el order, and the non-collinear double-$\mathbf{q}$ structure is a peculiar ``anti-collinear'' combination of the sublattices. In this view, while the sum of bilinear interactions between the sublattices vanishes, just like in the classic example of antiferromagnetic $J_2$-dominated square-lattice model \cite{HenleyPRL1989}, a non-collinear arrangement can be favored by higher-order interactions \cite{ZhouPRB2022}. Although conceptually useful, this four-sublattice picture cannot explain the different magnitudes of the spins, and there is no guarantee that the third-nearest-neighbor interactions in NCSO dominate over the nearer-neighbor ones.

The zero-field magnetic structure of NCSO holds importance for several reasons. First, the structure is double-$\mathbf{q}$ (Fig.~\ref{fig2}) instead of triple-$\mathbf{q}$, likely due to the strong magnetic in-plane anisotropy of NCSO \cite{LiPRX2022}. The surprising robustness of the multi-$\mathbf{q}$ order, despite the system's seemingly unfavorable symmetry, suggests that the order is stabilized by favorable higher-order exchange interactions \cite{KrugerArxiv2022, PohlePRB2023}. Second, the ordered spins are parallel to the $ab$-plane, supporting an XXZ starting point for the interaction model. A Kitaev-type model, with principle axes pointing at an angle out-of-plane due to the local crystallographic environment \cite{ChunNatPhys2015}, would need a significant coincidence to form purely in-plane ordering. Third, an XXZ starting point (as opposed to Kitaev) aligns NCSO with other honeycomb cobaltates \cite{HalloranPNAS2023, WinterJOP2022, DasPRB2021, LiuPRB2023}. While this may initially appear disadvantageous for quantum spin liquid exploration, the proposed double-$\mathbf{q}$ magnetic structure exhibits significantly reduced classical moments on half of the sites, suggesting strong quantum fluctuations \cite{SachdevNatPhys2008}. These fluctuations likely stem from a close competition between ferro- and antiferromagnetic ordering tendencies \cite{LiuPRL2020,LiPRX2022}. As antiferromagnetic order dominates at zero field, external fields could potentially drive the system to a tipping point, where stronger quantum fluctuations and spin-liquid behaviors might emerge.

In conclusion, we report experimental evidence of easy-plane multi-$\mathbf{q}$ magnetic order in NCSO. Our findings highlight the importance of considering high-order spin interactions within an XXZ framework when modeling honeycomb cobaltates previously considered candidate Kitaev materials. Although our results may require revisiting existing theories, NCSO remains a promising system for investigating novel phenomena related to spin liquids.

\begin{acknowledgments}

We are grateful for experimental assistance by Zirong Ye, Qian Xiao, Xiquan Zheng and Yingying Peng, and for discussions with Wenjie Chen, Lukas Janssen, Wilhelm G. F. Kr\"{u}ger, Zhengxin Liu, Yuan Wan, Jiucai Wang, Weiliang Yao, and Yi Zhou. The work at Peking University was supported by the National Basic Research Program of China (Grant No. 2021YFA1401900) and the NSF of China (Grant Nos. 12061131004 and 11888101). The work at Brookhaven National Laboratory was supported by Office of Basic Energy Sciences (BES), Division of Materials Sciences and Engineering, U.S. Department of Energy (DOE), under contract DE-SC0012704. A portion of this research used resources at Spallation Neutron Source, a DOE Office of Science User Facility operated by the Oak Ridge National Laboratory. One of the neutron scattering experiments was performed at the MLF, J-PARC, Japan, under a user program (No. 2021B0158).

\end{acknowledgments}

\bibliography{NCSO_multiq_ref}

\begin{thebibliography}{56}%
\makeatletter
\providecommand \@ifxundefined [1]{%
 \@ifx{#1\undefined}
}%
\providecommand \@ifnum [1]{%
 \ifnum #1\expandafter \@firstoftwo
 \else \expandafter \@secondoftwo
 \fi
}%
\providecommand \@ifx [1]{%
 \ifx #1\expandafter \@firstoftwo
 \else \expandafter \@secondoftwo
 \fi
}%
\providecommand \natexlab [1]{#1}%
\providecommand \enquote  [1]{``#1''}%
\providecommand \bibnamefont  [1]{#1}%
\providecommand \bibfnamefont [1]{#1}%
\providecommand \citenamefont [1]{#1}%
\providecommand \href@noop [0]{\@secondoftwo}%
\providecommand \href [0]{\begingroup \@sanitize@url \@href}%
\providecommand \@href[1]{\@@startlink{#1}\@@href}%
\providecommand \@@href[1]{\endgroup#1\@@endlink}%
\providecommand \@sanitize@url [0]{\catcode `\\12\catcode `\$12\catcode
  `\&12\catcode `\#12\catcode `\^12\catcode `\_12\catcode `\%12\relax}%
\providecommand \@@startlink[1]{}%
\providecommand \@@endlink[0]{}%
\providecommand \url  [0]{\begingroup\@sanitize@url \@url }%
\providecommand \@url [1]{\endgroup\@href {#1}{\urlprefix }}%
\providecommand \urlprefix  [0]{URL }%
\providecommand \Eprint [0]{\href }%
\providecommand \doibase [0]{https://doi.org/}%
\providecommand \selectlanguage [0]{\@gobble}%
\providecommand \bibinfo  [0]{\@secondoftwo}%
\providecommand \bibfield  [0]{\@secondoftwo}%
\providecommand \translation [1]{[#1]}%
\providecommand \BibitemOpen [0]{}%
\providecommand \bibitemStop [0]{}%
\providecommand \bibitemNoStop [0]{.\EOS\space}%
\providecommand \EOS [0]{\spacefactor3000\relax}%
\providecommand \BibitemShut  [1]{\csname bibitem#1\endcsname}%
\let\auto@bib@innerbib\@empty
\bibitem [{\citenamefont {Balents}(2010)}]{BalentsNature2010}%
  \BibitemOpen
  \bibfield  {author} {\bibinfo {author} {\bibfnamefont {L.}~\bibnamefont
  {Balents}},\ }\bibfield  {title} {\bibinfo {title} {Spin liquids in
  frustrated magnets},\ }\href {https://doi.org/10.1038/nature08917} {\bibfield
   {journal} {\bibinfo  {journal} {Nature}\ }\textbf {\bibinfo {volume}
  {464}},\ \bibinfo {pages} {199} (\bibinfo {year} {2010})}\BibitemShut
  {NoStop}%
\bibitem [{\citenamefont {Zhou}\ \emph {et~al.}(2017)\citenamefont {Zhou},
  \citenamefont {Kanoda},\ and\ \citenamefont {Ng}}]{ZhouRMP2017}%
  \BibitemOpen
  \bibfield  {author} {\bibinfo {author} {\bibfnamefont {Y.}~\bibnamefont
  {Zhou}}, \bibinfo {author} {\bibfnamefont {K.}~\bibnamefont {Kanoda}},\ and\
  \bibinfo {author} {\bibfnamefont {T.-K.}\ \bibnamefont {Ng}},\ }\bibfield
  {title} {\bibinfo {title} {Quantum spin liquid states},\ }\href
  {https://doi.org/10.1103/RevModPhys.89.025003} {\bibfield  {journal}
  {\bibinfo  {journal} {Rev. Mod. Phys.}\ }\textbf {\bibinfo {volume} {89}},\
  \bibinfo {pages} {025003} (\bibinfo {year} {2017})}\BibitemShut {NoStop}%
\bibitem [{\citenamefont {Broholm}\ \emph {et~al.}(2020)\citenamefont
  {Broholm}, \citenamefont {Cava}, \citenamefont {Kivelson}, \citenamefont
  {Nocera}, \citenamefont {Norman},\ and\ \citenamefont
  {Senthil}}]{BroholmScience2020}%
  \BibitemOpen
  \bibfield  {author} {\bibinfo {author} {\bibfnamefont {C.}~\bibnamefont
  {Broholm}}, \bibinfo {author} {\bibfnamefont {R.}~\bibnamefont {Cava}},
  \bibinfo {author} {\bibfnamefont {S.}~\bibnamefont {Kivelson}}, \bibinfo
  {author} {\bibfnamefont {D.}~\bibnamefont {Nocera}}, \bibinfo {author}
  {\bibfnamefont {M.}~\bibnamefont {Norman}},\ and\ \bibinfo {author}
  {\bibfnamefont {T.}~\bibnamefont {Senthil}},\ }\bibfield  {title} {\bibinfo
  {title} {Quantum spin liquids},\ }\href
  {https://doi.org/10.1126/science.aay0668} {\bibfield  {journal} {\bibinfo
  {journal} {Science}\ }\textbf {\bibinfo {volume} {367}},\ \bibinfo {pages}
  {eaay0668} (\bibinfo {year} {2020})}\BibitemShut {NoStop}%
\bibitem [{\citenamefont {Kitaev}(2006)}]{KitaevAP2006}%
  \BibitemOpen
  \bibfield  {author} {\bibinfo {author} {\bibfnamefont {A.}~\bibnamefont
  {Kitaev}},\ }\bibfield  {title} {\bibinfo {title} {Anyons in an exactly
  solved model and beyond},\ }\href
  {https://doi.org/https://doi.org/10.1016/j.aop.2005.10.005} {\bibfield
  {journal} {\bibinfo  {journal} {Ann. Phys.}\ }\textbf {\bibinfo {volume}
  {321}},\ \bibinfo {pages} {2 } (\bibinfo {year} {2006})}\BibitemShut
  {NoStop}%
\bibitem [{\citenamefont {Takagi}\ \emph {et~al.}(2019)\citenamefont {Takagi},
  \citenamefont {Takayama}, \citenamefont {Jackeli}, \citenamefont
  {Khaliullin},\ and\ \citenamefont {Nagler}}]{TakagiNRP2019}%
  \BibitemOpen
  \bibfield  {author} {\bibinfo {author} {\bibfnamefont {H.}~\bibnamefont
  {Takagi}}, \bibinfo {author} {\bibfnamefont {T.}~\bibnamefont {Takayama}},
  \bibinfo {author} {\bibfnamefont {G.}~\bibnamefont {Jackeli}}, \bibinfo
  {author} {\bibfnamefont {G.}~\bibnamefont {Khaliullin}},\ and\ \bibinfo
  {author} {\bibfnamefont {S.~E.}\ \bibnamefont {Nagler}},\ }\bibfield  {title}
  {\bibinfo {title} {{Concept and realization of Kitaev quantum spin
  liquids}},\ }\href {https://doi.org/10.1038/s42254-019-0038-2} {\bibfield
  {journal} {\bibinfo  {journal} {Nat. Rev. Phys.}\ }\textbf {\bibinfo {volume}
  {1}},\ \bibinfo {pages} {264} (\bibinfo {year} {2019})}\BibitemShut {NoStop}%
\bibitem [{\citenamefont {Jackeli}\ and\ \citenamefont
  {Khaliullin}(2009)}]{JackeliPRL2009}%
  \BibitemOpen
  \bibfield  {author} {\bibinfo {author} {\bibfnamefont {G.}~\bibnamefont
  {Jackeli}}\ and\ \bibinfo {author} {\bibfnamefont {G.}~\bibnamefont
  {Khaliullin}},\ }\bibfield  {title} {\bibinfo {title} {{Mott Insulators in
  the Strong Spin-Orbit Coupling Limit: From Heisenberg to a Quantum Compass
  and Kitaev Models}},\ }\href {https://doi.org/10.1103/PhysRevLett.102.017205}
  {\bibfield  {journal} {\bibinfo  {journal} {Phys. Rev. Lett.}\ }\textbf
  {\bibinfo {volume} {102}},\ \bibinfo {pages} {017205} (\bibinfo {year}
  {2009})}\BibitemShut {NoStop}%
\bibitem [{\citenamefont {Chaloupka}\ \emph {et~al.}(2010)\citenamefont
  {Chaloupka}, \citenamefont {Jackeli},\ and\ \citenamefont
  {Khaliullin}}]{ChaloupkaPRL2010}%
  \BibitemOpen
  \bibfield  {author} {\bibinfo {author} {\bibfnamefont {J.}~\bibnamefont
  {Chaloupka}}, \bibinfo {author} {\bibfnamefont {G.}~\bibnamefont {Jackeli}},\
  and\ \bibinfo {author} {\bibfnamefont {G.}~\bibnamefont {Khaliullin}},\
  }\bibfield  {title} {\bibinfo {title} {{Kitaev-Heisenberg Model on a
  Honeycomb Lattice: Possible Exotic Phases in Iridium Oxides
  ${A}_{2}{\mathrm{IrO}}_{3}$}},\ }\href
  {https://doi.org/10.1103/PhysRevLett.105.027204} {\bibfield  {journal}
  {\bibinfo  {journal} {Phys. Rev. Lett.}\ }\textbf {\bibinfo {volume} {105}},\
  \bibinfo {pages} {027204} (\bibinfo {year} {2010})}\BibitemShut {NoStop}%
\bibitem [{\citenamefont {Plumb}\ \emph {et~al.}(2014)\citenamefont {Plumb},
  \citenamefont {Clancy}, \citenamefont {Sandilands}, \citenamefont {Shankar},
  \citenamefont {Hu}, \citenamefont {Burch}, \citenamefont {Kee},\ and\
  \citenamefont {Kim}}]{PlumbPRB2014}%
  \BibitemOpen
  \bibfield  {author} {\bibinfo {author} {\bibfnamefont {K.~W.}\ \bibnamefont
  {Plumb}}, \bibinfo {author} {\bibfnamefont {J.~P.}\ \bibnamefont {Clancy}},
  \bibinfo {author} {\bibfnamefont {L.~J.}\ \bibnamefont {Sandilands}},
  \bibinfo {author} {\bibfnamefont {V.~V.}\ \bibnamefont {Shankar}}, \bibinfo
  {author} {\bibfnamefont {Y.~F.}\ \bibnamefont {Hu}}, \bibinfo {author}
  {\bibfnamefont {K.~S.}\ \bibnamefont {Burch}}, \bibinfo {author}
  {\bibfnamefont {H.-Y.}\ \bibnamefont {Kee}},\ and\ \bibinfo {author}
  {\bibfnamefont {Y.-J.}\ \bibnamefont {Kim}},\ }\bibfield  {title} {\bibinfo
  {title} {{$\ensuremath{\alpha}\ensuremath{-}{\mathrm{RuCl}}_{3}$: A
  spin-orbit assisted Mott insulator on a honeycomb lattice}},\ }\href
  {https://doi.org/10.1103/PhysRevB.90.041112} {\bibfield  {journal} {\bibinfo
  {journal} {Phys. Rev. B}\ }\textbf {\bibinfo {volume} {90}},\ \bibinfo
  {pages} {041112} (\bibinfo {year} {2014})}\BibitemShut {NoStop}%
\bibitem [{\citenamefont {Motome}\ \emph {et~al.}(2020)\citenamefont {Motome},
  \citenamefont {Sano}, \citenamefont {Jang}, \citenamefont {Sugita},\ and\
  \citenamefont {Kato}}]{MotomeJPCM2020}%
  \BibitemOpen
  \bibfield  {author} {\bibinfo {author} {\bibfnamefont {Y.}~\bibnamefont
  {Motome}}, \bibinfo {author} {\bibfnamefont {R.}~\bibnamefont {Sano}},
  \bibinfo {author} {\bibfnamefont {S.}~\bibnamefont {Jang}}, \bibinfo {author}
  {\bibfnamefont {Y.}~\bibnamefont {Sugita}},\ and\ \bibinfo {author}
  {\bibfnamefont {Y.}~\bibnamefont {Kato}},\ }\bibfield  {title} {\bibinfo
  {title} {{Materials design of Kitaev spin liquids beyond the
  Jackeli{\textendash}Khaliullin mechanism}},\ }\href
  {https://doi.org/10.1088/1361-648x/ab8525} {\bibfield  {journal} {\bibinfo
  {journal} {Journal of Physics: Condensed Matter}\ }\textbf {\bibinfo {volume}
  {32}},\ \bibinfo {pages} {404001} (\bibinfo {year} {2020})}\BibitemShut
  {NoStop}%
\bibitem [{\citenamefont {Trebst}\ and\ \citenamefont
  {Hickey}(2022)}]{TrebstPRRSP2022}%
  \BibitemOpen
  \bibfield  {author} {\bibinfo {author} {\bibfnamefont {S.}~\bibnamefont
  {Trebst}}\ and\ \bibinfo {author} {\bibfnamefont {C.}~\bibnamefont
  {Hickey}},\ }\bibfield  {title} {\bibinfo {title} {Kitaev materials},\ }\href
  {https://doi.org/https://doi.org/10.1016/j.physrep.2021.11.003} {\bibfield
  {journal} {\bibinfo  {journal} {Physics Reports}\ }\textbf {\bibinfo {volume}
  {950}},\ \bibinfo {pages} {1} (\bibinfo {year} {2022})}\BibitemShut {NoStop}%
\bibitem [{\citenamefont {Liu}\ and\ \citenamefont
  {Khaliullin}(2018)}]{LiuPRB2018}%
  \BibitemOpen
  \bibfield  {author} {\bibinfo {author} {\bibfnamefont {H.}~\bibnamefont
  {Liu}}\ and\ \bibinfo {author} {\bibfnamefont {G.}~\bibnamefont
  {Khaliullin}},\ }\bibfield  {title} {\bibinfo {title} {{Pseudospin exchange
  interactions in ${d}^{7}$ cobalt compounds: Possible realization of the
  Kitaev model}},\ }\href {https://doi.org/10.1103/PhysRevB.97.014407}
  {\bibfield  {journal} {\bibinfo  {journal} {Phys. Rev. B}\ }\textbf {\bibinfo
  {volume} {97}},\ \bibinfo {pages} {014407} (\bibinfo {year}
  {2018})}\BibitemShut {NoStop}%
\bibitem [{\citenamefont {Sano}\ \emph {et~al.}(2018)\citenamefont {Sano},
  \citenamefont {Kato},\ and\ \citenamefont {Motome}}]{SanoPRB2018}%
  \BibitemOpen
  \bibfield  {author} {\bibinfo {author} {\bibfnamefont {R.}~\bibnamefont
  {Sano}}, \bibinfo {author} {\bibfnamefont {Y.}~\bibnamefont {Kato}},\ and\
  \bibinfo {author} {\bibfnamefont {Y.}~\bibnamefont {Motome}},\ }\bibfield
  {title} {\bibinfo {title} {{Kitaev-Heisenberg Hamiltonian for high-spin
  ${d}^{7}$ Mott insulators}},\ }\href
  {https://doi.org/10.1103/PhysRevB.97.014408} {\bibfield  {journal} {\bibinfo
  {journal} {Phys. Rev. B}\ }\textbf {\bibinfo {volume} {97}},\ \bibinfo
  {pages} {014408} (\bibinfo {year} {2018})}\BibitemShut {NoStop}%
\bibitem [{\citenamefont {Liu}\ \emph {et~al.}(2020)\citenamefont {Liu},
  \citenamefont {Chaloupka},\ and\ \citenamefont {Khaliullin}}]{LiuPRL2020}%
  \BibitemOpen
  \bibfield  {author} {\bibinfo {author} {\bibfnamefont {H.}~\bibnamefont
  {Liu}}, \bibinfo {author} {\bibfnamefont {J.}~\bibnamefont {Chaloupka}},\
  and\ \bibinfo {author} {\bibfnamefont {G.}~\bibnamefont {Khaliullin}},\
  }\bibfield  {title} {\bibinfo {title} {Kitaev spin liquid in $3d$ transition
  metal compounds},\ }\href {https://doi.org/10.1103/PhysRevLett.125.047201}
  {\bibfield  {journal} {\bibinfo  {journal} {Phys. Rev. Lett.}\ }\textbf
  {\bibinfo {volume} {125}},\ \bibinfo {pages} {047201} (\bibinfo {year}
  {2020})}\BibitemShut {NoStop}%
\bibitem [{\citenamefont {Kim}\ \emph {et~al.}(2021{\natexlab{a}})\citenamefont
  {Kim}, \citenamefont {Kim},\ and\ \citenamefont {Park}}]{Kim_2021}%
  \BibitemOpen
  \bibfield  {author} {\bibinfo {author} {\bibfnamefont {C.}~\bibnamefont
  {Kim}}, \bibinfo {author} {\bibfnamefont {H.-S.}\ \bibnamefont {Kim}},\ and\
  \bibinfo {author} {\bibfnamefont {J.-G.}\ \bibnamefont {Park}},\ }\bibfield
  {title} {\bibinfo {title} {{Spin-orbital entangled state and realization of
  Kitaev physics in 3$d$ cobalt compounds: a progress report}},\ }\href
  {https://doi.org/10.1088/1361-648x/ac2d5d} {\bibfield  {journal} {\bibinfo
  {journal} {Journal of Physics: Condensed Matter}\ }\textbf {\bibinfo {volume}
  {34}},\ \bibinfo {pages} {023001} (\bibinfo {year}
  {2021}{\natexlab{a}})}\BibitemShut {NoStop}%
\bibitem [{\citenamefont {Liu}(2021)}]{LiuIJMPB2021}%
  \BibitemOpen
  \bibfield  {author} {\bibinfo {author} {\bibfnamefont {H.}~\bibnamefont
  {Liu}},\ }\bibfield  {title} {\bibinfo {title} {{Towards Kitaev spin liquid
  in $3d$ transition metal compounds}},\ }\href
  {https://doi.org/10.1142/S0217979221300061} {\bibfield  {journal} {\bibinfo
  {journal} {Int. J. Mod. Phys. B}\ }\textbf {\bibinfo {volume} {35}},\
  \bibinfo {pages} {2130006} (\bibinfo {year} {2021})}\BibitemShut {NoStop}%
\bibitem [{\citenamefont {Xiao}\ \emph {et~al.}(2019)\citenamefont {Xiao},
  \citenamefont {Xia}, \citenamefont {Zhang}, \citenamefont {Yue},
  \citenamefont {Huang}, \citenamefont {Zhang}, \citenamefont {Yang},
  \citenamefont {Song}, \citenamefont {Wei}, \citenamefont {Deng},\ and\
  \citenamefont {Jiang}}]{XiaoCGD2019}%
  \BibitemOpen
  \bibfield  {author} {\bibinfo {author} {\bibfnamefont {G.}~\bibnamefont
  {Xiao}}, \bibinfo {author} {\bibfnamefont {Z.}~\bibnamefont {Xia}}, \bibinfo
  {author} {\bibfnamefont {W.}~\bibnamefont {Zhang}}, \bibinfo {author}
  {\bibfnamefont {X.}~\bibnamefont {Yue}}, \bibinfo {author} {\bibfnamefont
  {S.}~\bibnamefont {Huang}}, \bibinfo {author} {\bibfnamefont
  {X.}~\bibnamefont {Zhang}}, \bibinfo {author} {\bibfnamefont
  {F.}~\bibnamefont {Yang}}, \bibinfo {author} {\bibfnamefont {Y.}~\bibnamefont
  {Song}}, \bibinfo {author} {\bibfnamefont {M.}~\bibnamefont {Wei}}, \bibinfo
  {author} {\bibfnamefont {H.}~\bibnamefont {Deng}},\ and\ \bibinfo {author}
  {\bibfnamefont {D.}~\bibnamefont {Jiang}},\ }\bibfield  {title} {\bibinfo
  {title} {{Crystal growth and the magnetic properties of Na$_2$Co$_2$TeO$_6$
  with quasi-two-dimensional honeycomb lattice}},\ }\href
  {https://doi.org/10.1021/acs.cgd.8b01770} {\bibfield  {journal} {\bibinfo
  {journal} {Cryst. Growth Des.}\ }\textbf {\bibinfo {volume} {19}},\ \bibinfo
  {pages} {2658} (\bibinfo {year} {2019})}\BibitemShut {NoStop}%
\bibitem [{\citenamefont {Yan}\ \emph {et~al.}(2019)\citenamefont {Yan},
  \citenamefont {Okamoto}, \citenamefont {Wu}, \citenamefont {Zheng},
  \citenamefont {Zhou}, \citenamefont {Cao},\ and\ \citenamefont
  {McGuire}}]{YanPRM2019}%
  \BibitemOpen
  \bibfield  {author} {\bibinfo {author} {\bibfnamefont {J.-Q.}\ \bibnamefont
  {Yan}}, \bibinfo {author} {\bibfnamefont {S.}~\bibnamefont {Okamoto}},
  \bibinfo {author} {\bibfnamefont {Y.}~\bibnamefont {Wu}}, \bibinfo {author}
  {\bibfnamefont {Q.}~\bibnamefont {Zheng}}, \bibinfo {author} {\bibfnamefont
  {H.~D.}\ \bibnamefont {Zhou}}, \bibinfo {author} {\bibfnamefont {H.~B.}\
  \bibnamefont {Cao}},\ and\ \bibinfo {author} {\bibfnamefont {M.~A.}\
  \bibnamefont {McGuire}},\ }\bibfield  {title} {\bibinfo {title} {{Magnetic
  order in single crystals of
  ${\mathrm{Na}}_{3}{\mathrm{Co}}_{2}{\mathrm{SbO}}_{6}$ with a honeycomb
  arrangement of
  ${3\mathrm{d}}^{7}\phantom{\rule{0.28em}{0ex}}{\mathrm{Co}}^{2+}$ ions}},\
  }\href {https://doi.org/10.1103/PhysRevMaterials.3.074405} {\bibfield
  {journal} {\bibinfo  {journal} {Phys. Rev. Mater.}\ }\textbf {\bibinfo
  {volume} {3}},\ \bibinfo {pages} {074405} (\bibinfo {year}
  {2019})}\BibitemShut {NoStop}%
\bibitem [{\citenamefont {Yao}\ and\ \citenamefont {Li}(2020)}]{YaoPRB2020}%
  \BibitemOpen
  \bibfield  {author} {\bibinfo {author} {\bibfnamefont {W.}~\bibnamefont
  {Yao}}\ and\ \bibinfo {author} {\bibfnamefont {Y.}~\bibnamefont {Li}},\
  }\bibfield  {title} {\bibinfo {title} {{Ferrimagnetism and Anisotropic Phase
  Tunability by Magnetic Fields in
  ${\mathrm{Na}}_{2}{\mathrm{Co}}_{2}{\mathrm{TeO}}_{6}$}},\ }\href
  {https://doi.org/10.1103/PhysRevB.101.085120} {\bibfield  {journal} {\bibinfo
   {journal} {Phys. Rev. B}\ }\textbf {\bibinfo {volume} {101}},\ \bibinfo
  {pages} {085120} (\bibinfo {year} {2020})}\BibitemShut {NoStop}%
\bibitem [{\citenamefont {Zhong}\ \emph {et~al.}(2020)\citenamefont {Zhong},
  \citenamefont {Gao}, \citenamefont {Ong},\ and\ \citenamefont
  {Cava}}]{ZhongSA2020}%
  \BibitemOpen
  \bibfield  {author} {\bibinfo {author} {\bibfnamefont {R.}~\bibnamefont
  {Zhong}}, \bibinfo {author} {\bibfnamefont {T.}~\bibnamefont {Gao}}, \bibinfo
  {author} {\bibfnamefont {N.~P.}\ \bibnamefont {Ong}},\ and\ \bibinfo {author}
  {\bibfnamefont {R.~J.}\ \bibnamefont {Cava}},\ }\bibfield  {title} {\bibinfo
  {title} {Weak-field induced nonmagnetic state in a co-based honeycomb},\
  }\href {https://doi.org/10.1126/sciadv.aay6953} {\bibfield  {journal}
  {\bibinfo  {journal} {Science Advances}\ }\textbf {\bibinfo {volume} {6}},\
  \bibinfo {pages} {eaay6953} (\bibinfo {year} {2020})}\BibitemShut {NoStop}%
\bibitem [{\citenamefont {Halloran}\ \emph {et~al.}(2023)\citenamefont
  {Halloran}, \citenamefont {Desrochers}, \citenamefont {Zhang}, \citenamefont
  {Chen}, \citenamefont {Chern}, \citenamefont {Xu}, \citenamefont {Winn},
  \citenamefont {Graves-Brook}, \citenamefont {Stone}, \citenamefont
  {Kolesnikov}, \citenamefont {Qiu}, \citenamefont {Zhong}, \citenamefont
  {Cava}, \citenamefont {Kim},\ and\ \citenamefont
  {Broholm}}]{HalloranPNAS2023}%
  \BibitemOpen
  \bibfield  {author} {\bibinfo {author} {\bibfnamefont {T.}~\bibnamefont
  {Halloran}}, \bibinfo {author} {\bibfnamefont {F.}~\bibnamefont
  {Desrochers}}, \bibinfo {author} {\bibfnamefont {E.~Z.}\ \bibnamefont
  {Zhang}}, \bibinfo {author} {\bibfnamefont {T.}~\bibnamefont {Chen}},
  \bibinfo {author} {\bibfnamefont {L.~E.}\ \bibnamefont {Chern}}, \bibinfo
  {author} {\bibfnamefont {Z.}~\bibnamefont {Xu}}, \bibinfo {author}
  {\bibfnamefont {B.}~\bibnamefont {Winn}}, \bibinfo {author} {\bibfnamefont
  {M.}~\bibnamefont {Graves-Brook}}, \bibinfo {author} {\bibfnamefont {M.~B.}\
  \bibnamefont {Stone}}, \bibinfo {author} {\bibfnamefont {A.~I.}\ \bibnamefont
  {Kolesnikov}}, \bibinfo {author} {\bibfnamefont {Y.}~\bibnamefont {Qiu}},
  \bibinfo {author} {\bibfnamefont {R.}~\bibnamefont {Zhong}}, \bibinfo
  {author} {\bibfnamefont {R.}~\bibnamefont {Cava}}, \bibinfo {author}
  {\bibfnamefont {Y.~B.}\ \bibnamefont {Kim}},\ and\ \bibinfo {author}
  {\bibfnamefont {C.}~\bibnamefont {Broholm}},\ }\bibfield  {title} {\bibinfo
  {title} {{Geometrical frustration versus Kitaev interactions in
  BaCo$_2$(AsO$_4$)$_2$}},\ }\href {https://doi.org/10.1073/pnas.2215509119}
  {\bibfield  {journal} {\bibinfo  {journal} {Proceedings of the National
  Academy of Sciences}\ }\textbf {\bibinfo {volume} {120}},\ \bibinfo {pages}
  {e2215509119} (\bibinfo {year} {2023})}\BibitemShut {NoStop}%
\bibitem [{\citenamefont {Winter}(2022)}]{WinterJOP2022}%
  \BibitemOpen
  \bibfield  {author} {\bibinfo {author} {\bibfnamefont {S.~M.}\ \bibnamefont
  {Winter}},\ }\bibfield  {title} {\bibinfo {title} {Magnetic couplings in
  edge-sharing high-spin $d^7$ compounds},\ }\href
  {https://doi.org/10.1088/2515-7639/ac94f8} {\bibfield  {journal} {\bibinfo
  {journal} {Journal of Physics: Materials}\ }\textbf {\bibinfo {volume} {5}},\
  \bibinfo {pages} {045003} (\bibinfo {year} {2022})}\BibitemShut {NoStop}%
\bibitem [{\citenamefont {Das}\ \emph {et~al.}(2021)\citenamefont {Das},
  \citenamefont {Voleti}, \citenamefont {Saha-Dasgupta},\ and\ \citenamefont
  {Paramekanti}}]{DasPRB2021}%
  \BibitemOpen
  \bibfield  {author} {\bibinfo {author} {\bibfnamefont {S.}~\bibnamefont
  {Das}}, \bibinfo {author} {\bibfnamefont {S.}~\bibnamefont {Voleti}},
  \bibinfo {author} {\bibfnamefont {T.}~\bibnamefont {Saha-Dasgupta}},\ and\
  \bibinfo {author} {\bibfnamefont {A.}~\bibnamefont {Paramekanti}},\
  }\bibfield  {title} {\bibinfo {title} {{XY magnetism, Kitaev exchange, and
  long-range frustration in the ${J}_{\mathrm{eff}}=\frac{1}{2}$ honeycomb
  cobaltates}},\ }\href {https://doi.org/10.1103/PhysRevB.104.134425}
  {\bibfield  {journal} {\bibinfo  {journal} {Phys. Rev. B}\ }\textbf {\bibinfo
  {volume} {104}},\ \bibinfo {pages} {134425} (\bibinfo {year}
  {2021})}\BibitemShut {NoStop}%
\bibitem [{\citenamefont {Liu}\ and\ \citenamefont {Kee}(2023)}]{LiuPRB2023}%
  \BibitemOpen
  \bibfield  {author} {\bibinfo {author} {\bibfnamefont {X.}~\bibnamefont
  {Liu}}\ and\ \bibinfo {author} {\bibfnamefont {H.-Y.}\ \bibnamefont {Kee}},\
  }\bibfield  {title} {\bibinfo {title} {{Non-Kitaev versus Kitaev honeycomb
  cobaltates}},\ }\href {https://doi.org/10.1103/PhysRevB.107.054420}
  {\bibfield  {journal} {\bibinfo  {journal} {Phys. Rev. B}\ }\textbf {\bibinfo
  {volume} {107}},\ \bibinfo {pages} {054420} (\bibinfo {year}
  {2023})}\BibitemShut {NoStop}%
\bibitem [{\citenamefont {Pandey}\ and\ \citenamefont
  {Feng}(2022)}]{PandeyPRB2022}%
  \BibitemOpen
  \bibfield  {author} {\bibinfo {author} {\bibfnamefont {S.~K.}\ \bibnamefont
  {Pandey}}\ and\ \bibinfo {author} {\bibfnamefont {J.}~\bibnamefont {Feng}},\
  }\bibfield  {title} {\bibinfo {title} {{Spin interaction and magnetism in
  cobaltate Kitaev candidate materials: An ab initio and model Hamiltonian
  approach}},\ }\href {https://doi.org/10.1103/PhysRevB.106.174411} {\bibfield
  {journal} {\bibinfo  {journal} {Phys. Rev. B}\ }\textbf {\bibinfo {volume}
  {106}},\ \bibinfo {pages} {174411} (\bibinfo {year} {2022})}\BibitemShut
  {NoStop}%
\bibitem [{\citenamefont {Li}\ \emph {et~al.}(2022)\citenamefont {Li},
  \citenamefont {Gu}, \citenamefont {Chen}, \citenamefont {Garlea},
  \citenamefont {Iida}, \citenamefont {Kamazawa}, \citenamefont {Li},
  \citenamefont {Deng}, \citenamefont {Xiao}, \citenamefont {Zheng},
  \citenamefont {Ye}, \citenamefont {Peng}, \citenamefont {Zaliznyak},
  \citenamefont {Tranquada},\ and\ \citenamefont {Li}}]{LiPRX2022}%
  \BibitemOpen
  \bibfield  {author} {\bibinfo {author} {\bibfnamefont {X.}~\bibnamefont
  {Li}}, \bibinfo {author} {\bibfnamefont {Y.}~\bibnamefont {Gu}}, \bibinfo
  {author} {\bibfnamefont {Y.}~\bibnamefont {Chen}}, \bibinfo {author}
  {\bibfnamefont {V.~O.}\ \bibnamefont {Garlea}}, \bibinfo {author}
  {\bibfnamefont {K.}~\bibnamefont {Iida}}, \bibinfo {author} {\bibfnamefont
  {K.}~\bibnamefont {Kamazawa}}, \bibinfo {author} {\bibfnamefont
  {Y.}~\bibnamefont {Li}}, \bibinfo {author} {\bibfnamefont {G.}~\bibnamefont
  {Deng}}, \bibinfo {author} {\bibfnamefont {Q.}~\bibnamefont {Xiao}}, \bibinfo
  {author} {\bibfnamefont {X.}~\bibnamefont {Zheng}}, \bibinfo {author}
  {\bibfnamefont {Z.}~\bibnamefont {Ye}}, \bibinfo {author} {\bibfnamefont
  {Y.}~\bibnamefont {Peng}}, \bibinfo {author} {\bibfnamefont {I.~A.}\
  \bibnamefont {Zaliznyak}}, \bibinfo {author} {\bibfnamefont {J.~M.}\
  \bibnamefont {Tranquada}},\ and\ \bibinfo {author} {\bibfnamefont
  {Y.}~\bibnamefont {Li}},\ }\bibfield  {title} {\bibinfo {title} {{Giant
  Magnetic In-Plane Anisotropy and Competing Instabilities in
  ${\mathrm{Na}}_{3}{\mathrm{Co}}_{2}{\mathrm{SbO}}_{6}$}},\ }\href
  {https://doi.org/10.1103/PhysRevX.12.041024} {\bibfield  {journal} {\bibinfo
  {journal} {Phys. Rev. X}\ }\textbf {\bibinfo {volume} {12}},\ \bibinfo
  {pages} {041024} (\bibinfo {year} {2022})}\BibitemShut {NoStop}%
\bibitem [{\citenamefont {Zhang}\ \emph {et~al.}(2023)\citenamefont {Zhang},
  \citenamefont {McGuire}, \citenamefont {May}, \citenamefont {Chao},
  \citenamefont {Zheng}, \citenamefont {Chi}, \citenamefont {Sales},
  \citenamefont {Mandrus}, \citenamefont {Nagler}, \citenamefont {Miao},
  \citenamefont {Ye},\ and\ \citenamefont {Yan}}]{Zhang2023Arxiv}%
  \BibitemOpen
  \bibfield  {author} {\bibinfo {author} {\bibfnamefont {H.}~\bibnamefont
  {Zhang}}, \bibinfo {author} {\bibfnamefont {M.~A.}\ \bibnamefont {McGuire}},
  \bibinfo {author} {\bibfnamefont {A.~F.}\ \bibnamefont {May}}, \bibinfo
  {author} {\bibfnamefont {J.}~\bibnamefont {Chao}}, \bibinfo {author}
  {\bibfnamefont {Q.}~\bibnamefont {Zheng}}, \bibinfo {author} {\bibfnamefont
  {M.}~\bibnamefont {Chi}}, \bibinfo {author} {\bibfnamefont {B.~C.}\
  \bibnamefont {Sales}}, \bibinfo {author} {\bibfnamefont {D.~G.}\ \bibnamefont
  {Mandrus}}, \bibinfo {author} {\bibfnamefont {S.~E.}\ \bibnamefont {Nagler}},
  \bibinfo {author} {\bibfnamefont {H.}~\bibnamefont {Miao}}, \bibinfo {author}
  {\bibfnamefont {F.}~\bibnamefont {Ye}},\ and\ \bibinfo {author}
  {\bibfnamefont {J.}~\bibnamefont {Yan}},\ }\bibfield  {title} {\bibinfo
  {title} {{Stacking disorder and thermal transport properties of
  $\alpha$-RuCl$_3$}}\ }\href {https://doi.org/10.48550/arXiv.2303.03682}
  {10.48550/arXiv.2303.03682} (\bibinfo {year} {2023}),\ \Eprint
  {https://arxiv.org/abs/2303.03682} {2303.03682} \BibitemShut {NoStop}%
\bibitem [{\citenamefont {Dufault}\ \emph {et~al.}(2023)\citenamefont
  {Dufault}, \citenamefont {Bahrami}, \citenamefont {Streeter}, \citenamefont
  {Yao}, \citenamefont {Gonzalez}, \citenamefont {Zhang},\ and\ \citenamefont
  {Tafti}}]{DufaultArxiv2023}%
  \BibitemOpen
  \bibfield  {author} {\bibinfo {author} {\bibfnamefont {E.}~\bibnamefont
  {Dufault}}, \bibinfo {author} {\bibfnamefont {F.}~\bibnamefont {Bahrami}},
  \bibinfo {author} {\bibfnamefont {A.}~\bibnamefont {Streeter}}, \bibinfo
  {author} {\bibfnamefont {X.}~\bibnamefont {Yao}}, \bibinfo {author}
  {\bibfnamefont {E.}~\bibnamefont {Gonzalez}}, \bibinfo {author}
  {\bibfnamefont {Q.}~\bibnamefont {Zhang}},\ and\ \bibinfo {author}
  {\bibfnamefont {F.}~\bibnamefont {Tafti}},\ }\bibfield  {title} {\bibinfo
  {title} {{Introducing the monoclinic polymorph of the Kitaev magnet
  Na$_2$Co$_2$TeO$_6$}}\ }\href {https://doi.org/10.48550/arXiv.2305.10484}
  {10.48550/arXiv.2305.10484} (\bibinfo {year} {2023})\BibitemShut {NoStop}%
\bibitem [{\citenamefont {Ohhara}\ \emph {et~al.}(2016)\citenamefont {Ohhara},
  \citenamefont {Kiyanagi}, \citenamefont {Oikawa}, \citenamefont {Kaneko},
  \citenamefont {Kawasaki}, \citenamefont {Tamura}, \citenamefont {Nakao},
  \citenamefont {Hanashima}, \citenamefont {Munakata}, \citenamefont {Moyoshi},
  \citenamefont {Kuroda}, \citenamefont {Kimura}, \citenamefont {Sakakura},
  \citenamefont {Lee}, \citenamefont {Takahashi}, \citenamefont {Ohshima},
  \citenamefont {Kiyotani}, \citenamefont {Noda},\ and\ \citenamefont
  {Arai}}]{OhharaJAC2016}%
  \BibitemOpen
  \bibfield  {author} {\bibinfo {author} {\bibfnamefont {T.}~\bibnamefont
  {Ohhara}}, \bibinfo {author} {\bibfnamefont {R.}~\bibnamefont {Kiyanagi}},
  \bibinfo {author} {\bibfnamefont {K.}~\bibnamefont {Oikawa}}, \bibinfo
  {author} {\bibfnamefont {K.}~\bibnamefont {Kaneko}}, \bibinfo {author}
  {\bibfnamefont {T.}~\bibnamefont {Kawasaki}}, \bibinfo {author}
  {\bibfnamefont {I.}~\bibnamefont {Tamura}}, \bibinfo {author} {\bibfnamefont
  {A.}~\bibnamefont {Nakao}}, \bibinfo {author} {\bibfnamefont
  {T.}~\bibnamefont {Hanashima}}, \bibinfo {author} {\bibfnamefont
  {K.}~\bibnamefont {Munakata}}, \bibinfo {author} {\bibfnamefont
  {T.}~\bibnamefont {Moyoshi}}, \bibinfo {author} {\bibfnamefont
  {T.}~\bibnamefont {Kuroda}}, \bibinfo {author} {\bibfnamefont
  {H.}~\bibnamefont {Kimura}}, \bibinfo {author} {\bibfnamefont
  {T.}~\bibnamefont {Sakakura}}, \bibinfo {author} {\bibfnamefont {C.-H.}\
  \bibnamefont {Lee}}, \bibinfo {author} {\bibfnamefont {M.}~\bibnamefont
  {Takahashi}}, \bibinfo {author} {\bibfnamefont {K.}~\bibnamefont {Ohshima}},
  \bibinfo {author} {\bibfnamefont {T.}~\bibnamefont {Kiyotani}}, \bibinfo
  {author} {\bibfnamefont {Y.}~\bibnamefont {Noda}},\ and\ \bibinfo {author}
  {\bibfnamefont {M.}~\bibnamefont {Arai}},\ }\bibfield  {title} {\bibinfo
  {title} {{SENJU: a new time-of-flight single-crystal neutron diffractometer
  at J-PARC}},\ }\href
  {https://doi.org/https://doi.org/10.1107/S1600576715022943} {\bibfield
  {journal} {\bibinfo  {journal} {J. Appl. Cryst}\ }\textbf {\bibinfo {volume}
  {49}},\ \bibinfo {pages} {120} (\bibinfo {year} {2016})}\BibitemShut
  {NoStop}%
\bibitem [{\citenamefont {Cao}\ \emph {et~al.}(2016)\citenamefont {Cao},
  \citenamefont {Banerjee}, \citenamefont {Yan}, \citenamefont {Bridges},
  \citenamefont {Lumsden}, \citenamefont {Mandrus}, \citenamefont {Tennant},
  \citenamefont {Chakoumakos},\ and\ \citenamefont {Nagler}}]{CaoPRB2016}%
  \BibitemOpen
  \bibfield  {author} {\bibinfo {author} {\bibfnamefont {H.~B.}\ \bibnamefont
  {Cao}}, \bibinfo {author} {\bibfnamefont {A.}~\bibnamefont {Banerjee}},
  \bibinfo {author} {\bibfnamefont {J.-Q.}\ \bibnamefont {Yan}}, \bibinfo
  {author} {\bibfnamefont {C.~A.}\ \bibnamefont {Bridges}}, \bibinfo {author}
  {\bibfnamefont {M.~D.}\ \bibnamefont {Lumsden}}, \bibinfo {author}
  {\bibfnamefont {D.~G.}\ \bibnamefont {Mandrus}}, \bibinfo {author}
  {\bibfnamefont {D.~A.}\ \bibnamefont {Tennant}}, \bibinfo {author}
  {\bibfnamefont {B.~C.}\ \bibnamefont {Chakoumakos}},\ and\ \bibinfo {author}
  {\bibfnamefont {S.~E.}\ \bibnamefont {Nagler}},\ }\bibfield  {title}
  {\bibinfo {title} {{Low-temperature crystal and magnetic structure of
  $\ensuremath{\alpha}\ensuremath{-}{\mathrm{RuCl}}_{3}$}},\ }\href
  {https://doi.org/10.1103/PhysRevB.93.134423} {\bibfield  {journal} {\bibinfo
  {journal} {Phys. Rev. B}\ }\textbf {\bibinfo {volume} {93}},\ \bibinfo
  {pages} {134423} (\bibinfo {year} {2016})}\BibitemShut {NoStop}%
\bibitem [{\citenamefont {Liu}\ \emph {et~al.}(2011)\citenamefont {Liu},
  \citenamefont {Berlijn}, \citenamefont {Yin}, \citenamefont {Ku},
  \citenamefont {Tsvelik}, \citenamefont {Kim}, \citenamefont {Gretarsson},
  \citenamefont {Singh}, \citenamefont {Gegenwart},\ and\ \citenamefont
  {Hill}}]{LiuPRB2011}%
  \BibitemOpen
  \bibfield  {author} {\bibinfo {author} {\bibfnamefont {X.}~\bibnamefont
  {Liu}}, \bibinfo {author} {\bibfnamefont {T.}~\bibnamefont {Berlijn}},
  \bibinfo {author} {\bibfnamefont {W.-G.}\ \bibnamefont {Yin}}, \bibinfo
  {author} {\bibfnamefont {W.}~\bibnamefont {Ku}}, \bibinfo {author}
  {\bibfnamefont {A.}~\bibnamefont {Tsvelik}}, \bibinfo {author} {\bibfnamefont
  {Y.-J.}\ \bibnamefont {Kim}}, \bibinfo {author} {\bibfnamefont
  {H.}~\bibnamefont {Gretarsson}}, \bibinfo {author} {\bibfnamefont
  {Y.}~\bibnamefont {Singh}}, \bibinfo {author} {\bibfnamefont
  {P.}~\bibnamefont {Gegenwart}},\ and\ \bibinfo {author} {\bibfnamefont
  {J.~P.}\ \bibnamefont {Hill}},\ }\bibfield  {title} {\bibinfo {title}
  {Long-range magnetic ordering in na${}_{2}$iro${}_{3}$},\ }\href
  {https://doi.org/10.1103/PhysRevB.83.220403} {\bibfield  {journal} {\bibinfo
  {journal} {Phys. Rev. B}\ }\textbf {\bibinfo {volume} {83}},\ \bibinfo
  {pages} {220403} (\bibinfo {year} {2011})}\BibitemShut {NoStop}%
\bibitem [{\citenamefont {Lefran\ifmmode~\mbox{\c{c}}\else \c{c}\fi{}ois}\
  \emph {et~al.}(2016)\citenamefont {Lefran\ifmmode~\mbox{\c{c}}\else
  \c{c}\fi{}ois}, \citenamefont {Songvilay}, \citenamefont {Robert},
  \citenamefont {Nataf}, \citenamefont {Jordan}, \citenamefont {Chaix},
  \citenamefont {Colin}, \citenamefont {Lejay}, \citenamefont {Hadj-Azzem},
  \citenamefont {Ballou},\ and\ \citenamefont {Simonet}}]{LefrancoisPRB2016}%
  \BibitemOpen
  \bibfield  {author} {\bibinfo {author} {\bibfnamefont {E.}~\bibnamefont
  {Lefran\ifmmode~\mbox{\c{c}}\else \c{c}\fi{}ois}}, \bibinfo {author}
  {\bibfnamefont {M.}~\bibnamefont {Songvilay}}, \bibinfo {author}
  {\bibfnamefont {J.}~\bibnamefont {Robert}}, \bibinfo {author} {\bibfnamefont
  {G.}~\bibnamefont {Nataf}}, \bibinfo {author} {\bibfnamefont
  {E.}~\bibnamefont {Jordan}}, \bibinfo {author} {\bibfnamefont
  {L.}~\bibnamefont {Chaix}}, \bibinfo {author} {\bibfnamefont {C.~V.}\
  \bibnamefont {Colin}}, \bibinfo {author} {\bibfnamefont {P.}~\bibnamefont
  {Lejay}}, \bibinfo {author} {\bibfnamefont {A.}~\bibnamefont {Hadj-Azzem}},
  \bibinfo {author} {\bibfnamefont {R.}~\bibnamefont {Ballou}},\ and\ \bibinfo
  {author} {\bibfnamefont {V.}~\bibnamefont {Simonet}},\ }\bibfield  {title}
  {\bibinfo {title} {{Magnetic properties of the honeycomb oxide
  ${\mathrm{Na}}_{2}{\mathrm{Co}}_{2}{\mathrm{TeO}}_{6}$}},\ }\href
  {https://doi.org/10.1103/PhysRevB.94.214416} {\bibfield  {journal} {\bibinfo
  {journal} {Phys. Rev. B}\ }\textbf {\bibinfo {volume} {94}},\ \bibinfo
  {pages} {214416} (\bibinfo {year} {2016})}\BibitemShut {NoStop}%
\bibitem [{\citenamefont {Bera}\ \emph {et~al.}(2017)\citenamefont {Bera},
  \citenamefont {Yusuf}, \citenamefont {Kumar},\ and\ \citenamefont
  {Ritter}}]{BeraPRB2017}%
  \BibitemOpen
  \bibfield  {author} {\bibinfo {author} {\bibfnamefont {A.~K.}\ \bibnamefont
  {Bera}}, \bibinfo {author} {\bibfnamefont {S.~M.}\ \bibnamefont {Yusuf}},
  \bibinfo {author} {\bibfnamefont {A.}~\bibnamefont {Kumar}},\ and\ \bibinfo
  {author} {\bibfnamefont {C.}~\bibnamefont {Ritter}},\ }\bibfield  {title}
  {\bibinfo {title} {Zigzag antiferromagnetic ground state with anisotropic
  correlation lengths in the quasi-two-dimensional honeycomb lattice compound
  $\mathrm{N}{\mathrm{a}}_{2}\mathrm{C}{\mathrm{o}}_{2}\mathrm{Te}{\mathrm{o}}_{6}$},\
  }\href {https://doi.org/10.1103/PhysRevB.95.094424} {\bibfield  {journal}
  {\bibinfo  {journal} {Phys. Rev. B}\ }\textbf {\bibinfo {volume} {95}},\
  \bibinfo {pages} {094424} (\bibinfo {year} {2017})}\BibitemShut {NoStop}%
\bibitem [{\citenamefont {Chen}\ \emph {et~al.}(2021)\citenamefont {Chen},
  \citenamefont {Li}, \citenamefont {Hu}, \citenamefont {Hu}, \citenamefont
  {Yue}, \citenamefont {Sutarto}, \citenamefont {He}, \citenamefont {Iida},
  \citenamefont {Kamazawa}, \citenamefont {Yu}, \citenamefont {Lin},\ and\
  \citenamefont {Li}}]{ChenPRB2021}%
  \BibitemOpen
  \bibfield  {author} {\bibinfo {author} {\bibfnamefont {W.}~\bibnamefont
  {Chen}}, \bibinfo {author} {\bibfnamefont {X.}~\bibnamefont {Li}}, \bibinfo
  {author} {\bibfnamefont {Z.}~\bibnamefont {Hu}}, \bibinfo {author}
  {\bibfnamefont {Z.}~\bibnamefont {Hu}}, \bibinfo {author} {\bibfnamefont
  {L.}~\bibnamefont {Yue}}, \bibinfo {author} {\bibfnamefont {R.}~\bibnamefont
  {Sutarto}}, \bibinfo {author} {\bibfnamefont {F.}~\bibnamefont {He}},
  \bibinfo {author} {\bibfnamefont {K.}~\bibnamefont {Iida}}, \bibinfo {author}
  {\bibfnamefont {K.}~\bibnamefont {Kamazawa}}, \bibinfo {author}
  {\bibfnamefont {W.}~\bibnamefont {Yu}}, \bibinfo {author} {\bibfnamefont
  {X.}~\bibnamefont {Lin}},\ and\ \bibinfo {author} {\bibfnamefont
  {Y.}~\bibnamefont {Li}},\ }\bibfield  {title} {\bibinfo {title} {{Spin-orbit
  phase behavior of ${\mathrm{Na}}_{2}{\mathrm{Co}}_{2}{\mathrm{TeO}}_{6}$ at
  low temperatures}},\ }\href {https://doi.org/10.1103/PhysRevB.103.L180404}
  {\bibfield  {journal} {\bibinfo  {journal} {Phys. Rev. B}\ }\textbf {\bibinfo
  {volume} {103}},\ \bibinfo {pages} {L180404} (\bibinfo {year}
  {2021})}\BibitemShut {NoStop}%
\bibitem [{\citenamefont {Kr\"{u}ger}\ \emph {et~al.}(2022)\citenamefont
  {Kr\"{u}ger}, \citenamefont {Chen}, \citenamefont {Jin}, \citenamefont {Li},\
  and\ \citenamefont {Janssen}}]{KrugerArxiv2022}%
  \BibitemOpen
  \bibfield  {author} {\bibinfo {author} {\bibfnamefont {W.~G.~F.}\
  \bibnamefont {Kr\"{u}ger}}, \bibinfo {author} {\bibfnamefont
  {W.}~\bibnamefont {Chen}}, \bibinfo {author} {\bibfnamefont {X.}~\bibnamefont
  {Jin}}, \bibinfo {author} {\bibfnamefont {Y.}~\bibnamefont {Li}},\ and\
  \bibinfo {author} {\bibfnamefont {L.}~\bibnamefont {Janssen}},\ }\bibfield
  {title} {\bibinfo {title} {{Triple-Q order in Na$_2$Co$_2$TeO$_6$ from
  proximity to hidden-SU(2)-symmetric point}}\ }\href
  {https://doi.org/10.48550/arXiv.2211.16957} {10.48550/arXiv.2211.16957}
  (\bibinfo {year} {2022})\BibitemShut {NoStop}%
\bibitem [{\citenamefont {Yao}\ \emph {et~al.}(2022)\citenamefont {Yao},
  \citenamefont {Iida}, \citenamefont {Kamazawa},\ and\ \citenamefont
  {Li}}]{YaoPRL2022}%
  \BibitemOpen
  \bibfield  {author} {\bibinfo {author} {\bibfnamefont {W.}~\bibnamefont
  {Yao}}, \bibinfo {author} {\bibfnamefont {K.}~\bibnamefont {Iida}}, \bibinfo
  {author} {\bibfnamefont {K.}~\bibnamefont {Kamazawa}},\ and\ \bibinfo
  {author} {\bibfnamefont {Y.}~\bibnamefont {Li}},\ }\bibfield  {title}
  {\bibinfo {title} {{Excitations in the Ordered and Paramagnetic States of
  Honeycomb Magnet ${\mathrm{Na}}_{2}{\mathrm{Co}}_{2}{\mathrm{TeO}}_{6}$}},\
  }\href {https://doi.org/10.1103/PhysRevLett.129.147202} {\bibfield  {journal}
  {\bibinfo  {journal} {Phys. Rev. Lett.}\ }\textbf {\bibinfo {volume} {129}},\
  \bibinfo {pages} {147202} (\bibinfo {year} {2022})}\BibitemShut {NoStop}%
\bibitem [{\citenamefont {Yao}\ \emph {et~al.}(2023)\citenamefont {Yao},
  \citenamefont {Zhao}, \citenamefont {Qiu}, \citenamefont {Balz},
  \citenamefont {Stewart}, \citenamefont {Lynn},\ and\ \citenamefont
  {Li}}]{YaoArxiv2022}%
  \BibitemOpen
  \bibfield  {author} {\bibinfo {author} {\bibfnamefont {W.}~\bibnamefont
  {Yao}}, \bibinfo {author} {\bibfnamefont {Y.}~\bibnamefont {Zhao}}, \bibinfo
  {author} {\bibfnamefont {Y.}~\bibnamefont {Qiu}}, \bibinfo {author}
  {\bibfnamefont {C.}~\bibnamefont {Balz}}, \bibinfo {author} {\bibfnamefont
  {J.~R.}\ \bibnamefont {Stewart}}, \bibinfo {author} {\bibfnamefont {J.~W.}\
  \bibnamefont {Lynn}},\ and\ \bibinfo {author} {\bibfnamefont
  {Y.}~\bibnamefont {Li}},\ }\bibfield  {title} {\bibinfo {title} {{Magnetic
  ground state of the Kitaev
  ${\mathrm{Na}}_{2}{\mathrm{Co}}_{2}{\mathrm{TeO}}_{6}$ spin liquid
  candidate}},\ }\href {https://doi.org/10.1103/PhysRevResearch.5.L022045}
  {\bibfield  {journal} {\bibinfo  {journal} {Phys. Rev. Res.}\ }\textbf
  {\bibinfo {volume} {5}},\ \bibinfo {pages} {L022045} (\bibinfo {year}
  {2023})}\BibitemShut {NoStop}%
\bibitem [{\citenamefont {Lee}\ \emph {et~al.}(2021)\citenamefont {Lee},
  \citenamefont {Lee}, \citenamefont {Choi}, \citenamefont {Jang},
  \citenamefont {Kalaivanan}, \citenamefont {Sankar},\ and\ \citenamefont
  {Choi}}]{LeePRB2021}%
  \BibitemOpen
  \bibfield  {author} {\bibinfo {author} {\bibfnamefont {C.~H.}\ \bibnamefont
  {Lee}}, \bibinfo {author} {\bibfnamefont {S.}~\bibnamefont {Lee}}, \bibinfo
  {author} {\bibfnamefont {Y.~S.}\ \bibnamefont {Choi}}, \bibinfo {author}
  {\bibfnamefont {Z.~H.}\ \bibnamefont {Jang}}, \bibinfo {author}
  {\bibfnamefont {R.}~\bibnamefont {Kalaivanan}}, \bibinfo {author}
  {\bibfnamefont {R.}~\bibnamefont {Sankar}},\ and\ \bibinfo {author}
  {\bibfnamefont {K.-Y.}\ \bibnamefont {Choi}},\ }\bibfield  {title} {\bibinfo
  {title} {{Multistage development of anisotropic magnetic correlations in the
  Co-based honeycomb lattice
  ${\mathrm{Na}}_{2}{\mathrm{Co}}_{2}{\mathrm{TeO}}_{6}$}},\ }\href
  {https://doi.org/10.1103/PhysRevB.103.214447} {\bibfield  {journal} {\bibinfo
   {journal} {Phys. Rev. B}\ }\textbf {\bibinfo {volume} {103}},\ \bibinfo
  {pages} {214447} (\bibinfo {year} {2021})}\BibitemShut {NoStop}%
\bibitem [{\citenamefont {Kikuchi}\ \emph {et~al.}(2022)\citenamefont
  {Kikuchi}, \citenamefont {Kamoda}, \citenamefont {Mera}, \citenamefont
  {Takahashi}, \citenamefont {Okumura},\ and\ \citenamefont
  {Yasui}}]{KikuchiArxiv2022}%
  \BibitemOpen
  \bibfield  {author} {\bibinfo {author} {\bibfnamefont {J.}~\bibnamefont
  {Kikuchi}}, \bibinfo {author} {\bibfnamefont {T.}~\bibnamefont {Kamoda}},
  \bibinfo {author} {\bibfnamefont {N.}~\bibnamefont {Mera}}, \bibinfo {author}
  {\bibfnamefont {Y.}~\bibnamefont {Takahashi}}, \bibinfo {author}
  {\bibfnamefont {K.}~\bibnamefont {Okumura}},\ and\ \bibinfo {author}
  {\bibfnamefont {Y.}~\bibnamefont {Yasui}},\ }\bibfield  {title} {\bibinfo
  {title} {{Field evolution of magnetic phases and spin dynamics in the
  honeycomb lattice magnet Na$_2$Co$_2$TeO$_6$: $^{23}$Na NMR study}},\ }\href
  {https://doi.org/10.48550/arXiv.2206.05409} {\bibfield  {journal} {\bibinfo
  {journal} {arXiv preprint arXiv:2206.05409}\ } (\bibinfo {year}
  {2022})}\BibitemShut {NoStop}%
\bibitem [{\citenamefont {Lin}\ \emph {et~al.}(2021)\citenamefont {Lin},
  \citenamefont {Jeong}, \citenamefont {Kim}, \citenamefont {Wang},
  \citenamefont {Huang}, \citenamefont {Masuda}, \citenamefont {Asai},
  \citenamefont {Itoh}, \citenamefont {G{\"u}nther}, \citenamefont {Russina}
  \emph {et~al.}}]{LinNC2021}%
  \BibitemOpen
  \bibfield  {author} {\bibinfo {author} {\bibfnamefont {G.}~\bibnamefont
  {Lin}}, \bibinfo {author} {\bibfnamefont {J.}~\bibnamefont {Jeong}}, \bibinfo
  {author} {\bibfnamefont {C.}~\bibnamefont {Kim}}, \bibinfo {author}
  {\bibfnamefont {Y.}~\bibnamefont {Wang}}, \bibinfo {author} {\bibfnamefont
  {Q.}~\bibnamefont {Huang}}, \bibinfo {author} {\bibfnamefont
  {T.}~\bibnamefont {Masuda}}, \bibinfo {author} {\bibfnamefont
  {S.}~\bibnamefont {Asai}}, \bibinfo {author} {\bibfnamefont {S.}~\bibnamefont
  {Itoh}}, \bibinfo {author} {\bibfnamefont {G.}~\bibnamefont {G{\"u}nther}},
  \bibinfo {author} {\bibfnamefont {M.}~\bibnamefont {Russina}}, \emph
  {et~al.},\ }\bibfield  {title} {\bibinfo {title} {{Field-induced quantum spin
  disordered state in spin-1/2 honeycomb magnet Na$_2$Co$_2$TeO$_6$}},\ }\href
  {https://doi.org/10.1038/s41467-021-25567-7} {\bibfield  {journal} {\bibinfo
  {journal} {Nature communications}\ }\textbf {\bibinfo {volume} {12}},\
  \bibinfo {pages} {5559} (\bibinfo {year} {2021})}\BibitemShut {NoStop}%
\bibitem [{\citenamefont {Songvilay}\ \emph {et~al.}(2020)\citenamefont
  {Songvilay}, \citenamefont {Robert}, \citenamefont {Petit}, \citenamefont
  {Rodriguez-Rivera}, \citenamefont {Ratcliff}, \citenamefont {Damay},
  \citenamefont {Bal\'edent}, \citenamefont {Jim\'enez-Ruiz}, \citenamefont
  {Lejay}, \citenamefont {Pachoud}, \citenamefont {Hadj-Azzem}, \citenamefont
  {Simonet},\ and\ \citenamefont {Stock}}]{SongvilayPRB2020}%
  \BibitemOpen
  \bibfield  {author} {\bibinfo {author} {\bibfnamefont {M.}~\bibnamefont
  {Songvilay}}, \bibinfo {author} {\bibfnamefont {J.}~\bibnamefont {Robert}},
  \bibinfo {author} {\bibfnamefont {S.}~\bibnamefont {Petit}}, \bibinfo
  {author} {\bibfnamefont {J.~A.}\ \bibnamefont {Rodriguez-Rivera}}, \bibinfo
  {author} {\bibfnamefont {W.~D.}\ \bibnamefont {Ratcliff}}, \bibinfo {author}
  {\bibfnamefont {F.}~\bibnamefont {Damay}}, \bibinfo {author} {\bibfnamefont
  {V.}~\bibnamefont {Bal\'edent}}, \bibinfo {author} {\bibfnamefont
  {M.}~\bibnamefont {Jim\'enez-Ruiz}}, \bibinfo {author} {\bibfnamefont
  {P.}~\bibnamefont {Lejay}}, \bibinfo {author} {\bibfnamefont
  {E.}~\bibnamefont {Pachoud}}, \bibinfo {author} {\bibfnamefont
  {A.}~\bibnamefont {Hadj-Azzem}}, \bibinfo {author} {\bibfnamefont
  {V.}~\bibnamefont {Simonet}},\ and\ \bibinfo {author} {\bibfnamefont
  {C.}~\bibnamefont {Stock}},\ }\bibfield  {title} {\bibinfo {title} {{Kitaev
  interactions in the Co honeycomb antiferromagnets
  ${\mathrm{Na}}_{3}{\mathrm{Co}}_{2}{\mathrm{SbO}}_{6}$ and
  ${\mathrm{Na}}_{2}{\mathrm{Co}}_{2}{\mathrm{TeO}}_{6}$}},\ }\href
  {https://doi.org/10.1103/PhysRevB.102.224429} {\bibfield  {journal} {\bibinfo
   {journal} {Phys. Rev. B}\ }\textbf {\bibinfo {volume} {102}},\ \bibinfo
  {pages} {224429} (\bibinfo {year} {2020})}\BibitemShut {NoStop}%
\bibitem [{\citenamefont {Kim}\ \emph {et~al.}(2021{\natexlab{b}})\citenamefont
  {Kim}, \citenamefont {Jeong}, \citenamefont {Lin}, \citenamefont {Park},
  \citenamefont {Masuda}, \citenamefont {Asai}, \citenamefont {Itoh},
  \citenamefont {Kim}, \citenamefont {Zhou}, \citenamefont {Ma},\ and\
  \citenamefont {Park}}]{KimJOP2021}%
  \BibitemOpen
  \bibfield  {author} {\bibinfo {author} {\bibfnamefont {C.}~\bibnamefont
  {Kim}}, \bibinfo {author} {\bibfnamefont {J.}~\bibnamefont {Jeong}}, \bibinfo
  {author} {\bibfnamefont {G.}~\bibnamefont {Lin}}, \bibinfo {author}
  {\bibfnamefont {P.}~\bibnamefont {Park}}, \bibinfo {author} {\bibfnamefont
  {T.}~\bibnamefont {Masuda}}, \bibinfo {author} {\bibfnamefont
  {S.}~\bibnamefont {Asai}}, \bibinfo {author} {\bibfnamefont {S.}~\bibnamefont
  {Itoh}}, \bibinfo {author} {\bibfnamefont {H.-S.}\ \bibnamefont {Kim}},
  \bibinfo {author} {\bibfnamefont {H.}~\bibnamefont {Zhou}}, \bibinfo {author}
  {\bibfnamefont {J.}~\bibnamefont {Ma}},\ and\ \bibinfo {author}
  {\bibfnamefont {J.-G.}\ \bibnamefont {Park}},\ }\bibfield  {title} {\bibinfo
  {title} {{Antiferromagnetic Kitaev interaction inJeff = 1/2 cobalt honeycomb
  materials Na$_3$Co$_2$SbO$_6$ and Na$_2$Co$_2$TeO$_6$}},\ }\href
  {https://doi.org/10.1088/1361-648X/ac2644} {\bibfield  {journal} {\bibinfo
  {journal} {Journal of Physics: Condensed Matter}\ }\textbf {\bibinfo {volume}
  {34}},\ \bibinfo {pages} {045802} (\bibinfo {year}
  {2021}{\natexlab{b}})}\BibitemShut {NoStop}%
\bibitem [{\citenamefont {Janssen}\ \emph {et~al.}(2016)\citenamefont
  {Janssen}, \citenamefont {Andrade},\ and\ \citenamefont
  {Vojta}}]{JanssenPRL2016}%
  \BibitemOpen
  \bibfield  {author} {\bibinfo {author} {\bibfnamefont {L.}~\bibnamefont
  {Janssen}}, \bibinfo {author} {\bibfnamefont {E.~C.}\ \bibnamefont
  {Andrade}},\ and\ \bibinfo {author} {\bibfnamefont {M.}~\bibnamefont
  {Vojta}},\ }\bibfield  {title} {\bibinfo {title} {{Honeycomb-Lattice
  Heisenberg-Kitaev Model in a Magnetic Field: Spin Canting, Metamagnetism, and
  Vortex Crystals}},\ }\href {https://doi.org/10.1103/PhysRevLett.117.277202}
  {\bibfield  {journal} {\bibinfo  {journal} {Phys. Rev. Lett.}\ }\textbf
  {\bibinfo {volume} {117}},\ \bibinfo {pages} {277202} (\bibinfo {year}
  {2016})}\BibitemShut {NoStop}%
\bibitem [{\citenamefont {Pohle}\ \emph {et~al.}(2023)\citenamefont {Pohle},
  \citenamefont {Shannon},\ and\ \citenamefont {Motome}}]{PohlePRB2023}%
  \BibitemOpen
  \bibfield  {author} {\bibinfo {author} {\bibfnamefont {R.}~\bibnamefont
  {Pohle}}, \bibinfo {author} {\bibfnamefont {N.}~\bibnamefont {Shannon}},\
  and\ \bibinfo {author} {\bibfnamefont {Y.}~\bibnamefont {Motome}},\
  }\bibfield  {title} {\bibinfo {title} {{Spin nematics meet spin liquids:
  Exotic quantum phases in the spin-1 bilinear-biquadratic model with Kitaev
  interactions}},\ }\href {https://doi.org/10.1103/PhysRevB.107.L140403}
  {\bibfield  {journal} {\bibinfo  {journal} {Phys. Rev. B}\ }\textbf {\bibinfo
  {volume} {107}},\ \bibinfo {pages} {L140403} (\bibinfo {year}
  {2023})}\BibitemShut {NoStop}%
\bibitem [{\citenamefont {Wang}\ and\ \citenamefont
  {Liu}(2023)}]{WangarXiv2023}%
  \BibitemOpen
  \bibfield  {author} {\bibinfo {author} {\bibfnamefont {J.}~\bibnamefont
  {Wang}}\ and\ \bibinfo {author} {\bibfnamefont {Z.}~\bibnamefont {Liu}},\
  }\bibfield  {title} {\bibinfo {title} {{Effect of ring-exchange interactions
  in the extended Kitaev honeycomb model}}\ }\href
  {https://doi.org/10.48550/arXiv.2305.03258} {10.48550/arXiv.2305.03258}
  (\bibinfo {year} {2023})\BibitemShut {NoStop}%
\bibitem [{\citenamefont {Johnson}\ \emph {et~al.}(2015)\citenamefont
  {Johnson}, \citenamefont {Williams}, \citenamefont {Haghighirad},
  \citenamefont {Singleton}, \citenamefont {Zapf}, \citenamefont {Manuel},
  \citenamefont {Mazin}, \citenamefont {Li}, \citenamefont {Jeschke},
  \citenamefont {Valent\'{\i}},\ and\ \citenamefont {Coldea}}]{JohnsonPRB2015}%
  \BibitemOpen
  \bibfield  {author} {\bibinfo {author} {\bibfnamefont {R.~D.}\ \bibnamefont
  {Johnson}}, \bibinfo {author} {\bibfnamefont {S.~C.}\ \bibnamefont
  {Williams}}, \bibinfo {author} {\bibfnamefont {A.~A.}\ \bibnamefont
  {Haghighirad}}, \bibinfo {author} {\bibfnamefont {J.}~\bibnamefont
  {Singleton}}, \bibinfo {author} {\bibfnamefont {V.}~\bibnamefont {Zapf}},
  \bibinfo {author} {\bibfnamefont {P.}~\bibnamefont {Manuel}}, \bibinfo
  {author} {\bibfnamefont {I.~I.}\ \bibnamefont {Mazin}}, \bibinfo {author}
  {\bibfnamefont {Y.}~\bibnamefont {Li}}, \bibinfo {author} {\bibfnamefont
  {H.~O.}\ \bibnamefont {Jeschke}}, \bibinfo {author} {\bibfnamefont
  {R.}~\bibnamefont {Valent\'{\i}}},\ and\ \bibinfo {author} {\bibfnamefont
  {R.}~\bibnamefont {Coldea}},\ }\bibfield  {title} {\bibinfo {title}
  {Monoclinic crystal structure of $\alpha$-\ch{RuCl_3} and the zigzag
  antiferromagnetic ground state},\ }\href
  {https://doi.org/10.1103/PhysRevB.92.235119} {\bibfield  {journal} {\bibinfo
  {journal} {Phys. Rev. B}\ }\textbf {\bibinfo {volume} {92}},\ \bibinfo
  {pages} {235119} (\bibinfo {year} {2015})}\BibitemShut {NoStop}%
\bibitem [{\citenamefont {Michels-Clark}\ \emph {et~al.}(2016)\citenamefont
  {Michels-Clark}, \citenamefont {Savici}, \citenamefont {Lynch}, \citenamefont
  {Wang},\ and\ \citenamefont {Hoffmann}}]{MichelsClarkJAC2016}%
  \BibitemOpen
  \bibfield  {author} {\bibinfo {author} {\bibfnamefont {T.~M.}\ \bibnamefont
  {Michels-Clark}}, \bibinfo {author} {\bibfnamefont {A.~T.}\ \bibnamefont
  {Savici}}, \bibinfo {author} {\bibfnamefont {V.~E.}\ \bibnamefont {Lynch}},
  \bibinfo {author} {\bibfnamefont {X.}~\bibnamefont {Wang}},\ and\ \bibinfo
  {author} {\bibfnamefont {C.~M.}\ \bibnamefont {Hoffmann}},\ }\bibfield
  {title} {\bibinfo {title} {{Expanding Lorentz and spectrum corrections to
  large volumes of reciprocal space for single-crystal time-of-flight neutron
  diffraction}},\ }\href {https://doi.org/10.1107/S1600576716001369} {\bibfield
   {journal} {\bibinfo  {journal} {Journal of Applied Crystallography}\
  }\textbf {\bibinfo {volume} {49}},\ \bibinfo {pages} {497} (\bibinfo {year}
  {2016})}\BibitemShut {NoStop}%
\bibitem [{SM()}]{SM}%
  \BibitemOpen
  \bibinfo {note} {See Supplemental Material for additional methods and data,
  which include additional Refs. \cite{WuJOI2016,BarryEPJ2015}.}\BibitemShut
  {Stop}%
\bibitem [{\citenamefont {Zaliznyak}\ \emph {et~al.}(2017)\citenamefont
  {Zaliznyak}, \citenamefont {Savici}, \citenamefont {Garlea}, \citenamefont
  {Winn}, \citenamefont {Filges}, \citenamefont {Schneeloch}, \citenamefont
  {Tranquada}, \citenamefont {Gu}, \citenamefont {Wang},\ and\ \citenamefont
  {Petrovic}}]{ZaliznyakIOP2017}%
  \BibitemOpen
  \bibfield  {author} {\bibinfo {author} {\bibfnamefont {I.~A.}\ \bibnamefont
  {Zaliznyak}}, \bibinfo {author} {\bibfnamefont {A.~T.}\ \bibnamefont
  {Savici}}, \bibinfo {author} {\bibfnamefont {V.~O.}\ \bibnamefont {Garlea}},
  \bibinfo {author} {\bibfnamefont {B.}~\bibnamefont {Winn}}, \bibinfo {author}
  {\bibfnamefont {U.}~\bibnamefont {Filges}}, \bibinfo {author} {\bibfnamefont
  {J.}~\bibnamefont {Schneeloch}}, \bibinfo {author} {\bibfnamefont {J.~M.}\
  \bibnamefont {Tranquada}}, \bibinfo {author} {\bibfnamefont {G.}~\bibnamefont
  {Gu}}, \bibinfo {author} {\bibfnamefont {A.}~\bibnamefont {Wang}},\ and\
  \bibinfo {author} {\bibfnamefont {C.}~\bibnamefont {Petrovic}},\ }\bibfield
  {title} {\bibinfo {title} {{Polarized neutron scattering on HYSPEC: the
  HYbrid SPECtrometer at SNS}},\ }\href
  {https://doi.org/10.1088/1742-6596/862/1/012030} {\bibfield  {journal}
  {\bibinfo  {journal} {Journal of Physics: Conference Series}\ }\textbf
  {\bibinfo {volume} {862}},\ \bibinfo {pages} {012030} (\bibinfo {year}
  {2017})}\BibitemShut {NoStop}%
\bibitem [{\citenamefont {Choi}\ \emph {et~al.}(2012)\citenamefont {Choi},
  \citenamefont {Coldea}, \citenamefont {Kolmogorov}, \citenamefont
  {Lancaster}, \citenamefont {Mazin}, \citenamefont {Blundell}, \citenamefont
  {Radaelli}, \citenamefont {Singh}, \citenamefont {Gegenwart}, \citenamefont
  {Choi}, \citenamefont {Cheong}, \citenamefont {Baker}, \citenamefont
  {Stock},\ and\ \citenamefont {Taylor}}]{ChoiPRL2012}%
  \BibitemOpen
  \bibfield  {author} {\bibinfo {author} {\bibfnamefont {S.~K.}\ \bibnamefont
  {Choi}}, \bibinfo {author} {\bibfnamefont {R.}~\bibnamefont {Coldea}},
  \bibinfo {author} {\bibfnamefont {A.~N.}\ \bibnamefont {Kolmogorov}},
  \bibinfo {author} {\bibfnamefont {T.}~\bibnamefont {Lancaster}}, \bibinfo
  {author} {\bibfnamefont {I.~I.}\ \bibnamefont {Mazin}}, \bibinfo {author}
  {\bibfnamefont {S.~J.}\ \bibnamefont {Blundell}}, \bibinfo {author}
  {\bibfnamefont {P.~G.}\ \bibnamefont {Radaelli}}, \bibinfo {author}
  {\bibfnamefont {Y.}~\bibnamefont {Singh}}, \bibinfo {author} {\bibfnamefont
  {P.}~\bibnamefont {Gegenwart}}, \bibinfo {author} {\bibfnamefont {K.~R.}\
  \bibnamefont {Choi}}, \bibinfo {author} {\bibfnamefont {S.-W.}\ \bibnamefont
  {Cheong}}, \bibinfo {author} {\bibfnamefont {P.~J.}\ \bibnamefont {Baker}},
  \bibinfo {author} {\bibfnamefont {C.}~\bibnamefont {Stock}},\ and\ \bibinfo
  {author} {\bibfnamefont {J.}~\bibnamefont {Taylor}},\ }\bibfield  {title}
  {\bibinfo {title} {{Spin Waves and Revised Crystal Structure of Honeycomb
  Iridate ${\mathrm{Na}}_{2}{\mathrm{IrO}}_{3}$}},\ }\href
  {https://doi.org/10.1103/PhysRevLett.108.127204} {\bibfield  {journal}
  {\bibinfo  {journal} {Phys. Rev. Lett.}\ }\textbf {\bibinfo {volume} {108}},\
  \bibinfo {pages} {127204} (\bibinfo {year} {2012})}\BibitemShut {NoStop}%
\bibitem [{\citenamefont {Maksimov}\ \emph {et~al.}(2022)\citenamefont
  {Maksimov}, \citenamefont {Ushakov}, \citenamefont {Pchelkina}, \citenamefont
  {Li}, \citenamefont {Winter},\ and\ \citenamefont
  {Streltsov}}]{MaksimovPRB2022}%
  \BibitemOpen
  \bibfield  {author} {\bibinfo {author} {\bibfnamefont {P.~A.}\ \bibnamefont
  {Maksimov}}, \bibinfo {author} {\bibfnamefont {A.~V.}\ \bibnamefont
  {Ushakov}}, \bibinfo {author} {\bibfnamefont {Z.~V.}\ \bibnamefont
  {Pchelkina}}, \bibinfo {author} {\bibfnamefont {Y.}~\bibnamefont {Li}},
  \bibinfo {author} {\bibfnamefont {S.~M.}\ \bibnamefont {Winter}},\ and\
  \bibinfo {author} {\bibfnamefont {S.~V.}\ \bibnamefont {Streltsov}},\
  }\bibfield  {title} {\bibinfo {title} {{Ab initio guided minimal model for
  the ``Kitaev'' material ${\mathrm{BaCo}}_{2}$(${\mathrm{AsO}}_{4}{)}_{2}$:
  Importance of direct hopping, third-neighbor exchange, and quantum
  fluctuations}},\ }\href {https://doi.org/10.1103/PhysRevB.106.165131}
  {\bibfield  {journal} {\bibinfo  {journal} {Phys. Rev. B}\ }\textbf {\bibinfo
  {volume} {106}},\ \bibinfo {pages} {165131} (\bibinfo {year}
  {2022})}\BibitemShut {NoStop}%
\bibitem [{\citenamefont {Henley}(1989)}]{HenleyPRL1989}%
  \BibitemOpen
  \bibfield  {author} {\bibinfo {author} {\bibfnamefont {C.~L.}\ \bibnamefont
  {Henley}},\ }\bibfield  {title} {\bibinfo {title} {Ordering due to disorder
  in a frustrated vector antiferromagnet},\ }\href
  {https://doi.org/10.1103/PhysRevLett.62.2056} {\bibfield  {journal} {\bibinfo
   {journal} {Phys. Rev. Lett.}\ }\textbf {\bibinfo {volume} {62}},\ \bibinfo
  {pages} {2056} (\bibinfo {year} {1989})}\BibitemShut {NoStop}%
\bibitem [{\citenamefont {Zhou}\ \emph {et~al.}(2022)\citenamefont {Zhou},
  \citenamefont {Quirion}, \citenamefont {Quilliam}, \citenamefont {Cao},
  \citenamefont {Ye}, \citenamefont {Stone}, \citenamefont {Huang},
  \citenamefont {Zhou}, \citenamefont {Cheng}, \citenamefont {Bai},
  \citenamefont {Mourigal}, \citenamefont {Wan},\ and\ \citenamefont
  {Dun}}]{ZhouPRB2022}%
  \BibitemOpen
  \bibfield  {author} {\bibinfo {author} {\bibfnamefont {J.}~\bibnamefont
  {Zhou}}, \bibinfo {author} {\bibfnamefont {G.}~\bibnamefont {Quirion}},
  \bibinfo {author} {\bibfnamefont {J.~A.}\ \bibnamefont {Quilliam}}, \bibinfo
  {author} {\bibfnamefont {H.}~\bibnamefont {Cao}}, \bibinfo {author}
  {\bibfnamefont {F.}~\bibnamefont {Ye}}, \bibinfo {author} {\bibfnamefont
  {M.~B.}\ \bibnamefont {Stone}}, \bibinfo {author} {\bibfnamefont
  {Q.}~\bibnamefont {Huang}}, \bibinfo {author} {\bibfnamefont
  {H.}~\bibnamefont {Zhou}}, \bibinfo {author} {\bibfnamefont {J.}~\bibnamefont
  {Cheng}}, \bibinfo {author} {\bibfnamefont {X.}~\bibnamefont {Bai}}, \bibinfo
  {author} {\bibfnamefont {M.}~\bibnamefont {Mourigal}}, \bibinfo {author}
  {\bibfnamefont {Y.}~\bibnamefont {Wan}},\ and\ \bibinfo {author}
  {\bibfnamefont {Z.}~\bibnamefont {Dun}},\ }\bibfield  {title} {\bibinfo
  {title} {{Anticollinear order and degeneracy lifting in square lattice
  antiferromagnet ${\mathrm{LaSrCrO}}_{4}$}},\ }\href
  {https://doi.org/10.1103/PhysRevB.105.L180411} {\bibfield  {journal}
  {\bibinfo  {journal} {Phys. Rev. B}\ }\textbf {\bibinfo {volume} {105}},\
  \bibinfo {pages} {L180411} (\bibinfo {year} {2022})}\BibitemShut {NoStop}%
\bibitem [{\citenamefont {Chun}\ \emph {et~al.}(2015)\citenamefont {Chun},
  \citenamefont {Kim}, \citenamefont {Kim}, \citenamefont {Zheng},
  \citenamefont {Stoumpos}, \citenamefont {Malliakas}, \citenamefont
  {Mitchell}, \citenamefont {Mehlawat}, \citenamefont {Singh}, \citenamefont
  {Choi}, \citenamefont {Gog}, \citenamefont {Al-Zein}, \citenamefont
  {Moretti~Sala}, \citenamefont {Krisch}, \citenamefont {Chaloupka},
  \citenamefont {Jackeli}, \citenamefont {Khaliullin},\ and\ \citenamefont
  {Kim}}]{ChunNatPhys2015}%
  \BibitemOpen
  \bibfield  {author} {\bibinfo {author} {\bibfnamefont {S.~H.}\ \bibnamefont
  {Chun}}, \bibinfo {author} {\bibfnamefont {J.-W.}\ \bibnamefont {Kim}},
  \bibinfo {author} {\bibfnamefont {J.}~\bibnamefont {Kim}}, \bibinfo {author}
  {\bibfnamefont {H.}~\bibnamefont {Zheng}}, \bibinfo {author} {\bibfnamefont
  {C.~C.}\ \bibnamefont {Stoumpos}}, \bibinfo {author} {\bibfnamefont {C.~D.}\
  \bibnamefont {Malliakas}}, \bibinfo {author} {\bibfnamefont {J.~F.}\
  \bibnamefont {Mitchell}}, \bibinfo {author} {\bibfnamefont {K.}~\bibnamefont
  {Mehlawat}}, \bibinfo {author} {\bibfnamefont {Y.}~\bibnamefont {Singh}},
  \bibinfo {author} {\bibfnamefont {Y.}~\bibnamefont {Choi}}, \bibinfo {author}
  {\bibfnamefont {T.}~\bibnamefont {Gog}}, \bibinfo {author} {\bibfnamefont
  {A.}~\bibnamefont {Al-Zein}}, \bibinfo {author} {\bibfnamefont
  {M.}~\bibnamefont {Moretti~Sala}}, \bibinfo {author} {\bibfnamefont
  {M.}~\bibnamefont {Krisch}}, \bibinfo {author} {\bibfnamefont
  {J.}~\bibnamefont {Chaloupka}}, \bibinfo {author} {\bibfnamefont
  {G.}~\bibnamefont {Jackeli}}, \bibinfo {author} {\bibfnamefont
  {G.}~\bibnamefont {Khaliullin}},\ and\ \bibinfo {author} {\bibfnamefont
  {B.~J.}\ \bibnamefont {Kim}},\ }\bibfield  {title} {\bibinfo {title} {{Direct
  evidence for dominant bond-directional interactions in a honeycomb lattice
  iridate Na$_2$IrO$_3$}},\ }\href {https://doi.org/10.1038/NPHYS3322}
  {\bibfield  {journal} {\bibinfo  {journal} {Nature Physics}\ }\textbf
  {\bibinfo {volume} {11}},\ \bibinfo {pages} {462} (\bibinfo {year}
  {2015})}\BibitemShut {NoStop}%
\bibitem [{\citenamefont {Sachdev}(2008)}]{SachdevNatPhys2008}%
  \BibitemOpen
  \bibfield  {author} {\bibinfo {author} {\bibfnamefont {S.}~\bibnamefont
  {Sachdev}},\ }\bibfield  {title} {\bibinfo {title} {{Quantum magnetism and
  criticality}},\ }\href {https://doi.org/10.1038/nphys894} {\bibfield
  {journal} {\bibinfo  {journal} {Nat. Phys.}\ }\textbf {\bibinfo {volume}
  {4}},\ \bibinfo {pages} {173} (\bibinfo {year} {2008})}\BibitemShut {NoStop}%
\bibitem [{\citenamefont {Wu}\ \emph {et~al.}(2016)\citenamefont {Wu},
  \citenamefont {Deng}, \citenamefont {Gardner}, \citenamefont {Vorderwisch},
  \citenamefont {Li}, \citenamefont {Yano}, \citenamefont {Peng},\ and\
  \citenamefont {Imamovic}}]{WuJOI2016}%
  \BibitemOpen
  \bibfield  {author} {\bibinfo {author} {\bibfnamefont {C.-M.}\ \bibnamefont
  {Wu}}, \bibinfo {author} {\bibfnamefont {G.}~\bibnamefont {Deng}}, \bibinfo
  {author} {\bibfnamefont {J.}~\bibnamefont {Gardner}}, \bibinfo {author}
  {\bibfnamefont {P.}~\bibnamefont {Vorderwisch}}, \bibinfo {author}
  {\bibfnamefont {W.-H.}\ \bibnamefont {Li}}, \bibinfo {author} {\bibfnamefont
  {S.}~\bibnamefont {Yano}}, \bibinfo {author} {\bibfnamefont {J.-C.}\
  \bibnamefont {Peng}},\ and\ \bibinfo {author} {\bibfnamefont
  {E.}~\bibnamefont {Imamovic}},\ }\bibfield  {title} {\bibinfo {title}
  {{SIKA}{\textemdash}the multiplexing cold-neutron triple-axis spectrometer at
  {ANSTO}},\ }\href {https://doi.org/10.1088/1748-0221/11/10/p10009} {\bibfield
   {journal} {\bibinfo  {journal} {Journal of Instrumentation}\ }\textbf
  {\bibinfo {volume} {11}}\bibinfo  {number} { (10)},\ \bibinfo {pages}
  {P10009}}\BibitemShut {NoStop}%
\bibitem [{\citenamefont {Winn}\ \emph {et~al.}(2015)\citenamefont {Winn},
  \citenamefont {Filges}, \citenamefont {Garlea}, \citenamefont {Graves-Brook},
  \citenamefont {Hagen}, \citenamefont {Jiang}, \citenamefont {Kenzelmann},
  \citenamefont {Passell}, \citenamefont {Shapiro}, \citenamefont {Tong},\ and\
  \citenamefont {Zaliznyak}}]{BarryEPJ2015}%
  \BibitemOpen
\bibfield  {number} {  }\bibfield  {author} {\bibinfo {author} {\bibfnamefont
  {B.}~\bibnamefont {Winn}}, \bibinfo {author} {\bibfnamefont {U.}~\bibnamefont
  {Filges}}, \bibinfo {author} {\bibfnamefont {V.~O.}\ \bibnamefont {Garlea}},
  \bibinfo {author} {\bibfnamefont {M.}~\bibnamefont {Graves-Brook}}, \bibinfo
  {author} {\bibfnamefont {M.}~\bibnamefont {Hagen}}, \bibinfo {author}
  {\bibfnamefont {C.}~\bibnamefont {Jiang}}, \bibinfo {author} {\bibfnamefont
  {M.}~\bibnamefont {Kenzelmann}}, \bibinfo {author} {\bibfnamefont
  {L.}~\bibnamefont {Passell}}, \bibinfo {author} {\bibfnamefont {S.~M.}\
  \bibnamefont {Shapiro}}, \bibinfo {author} {\bibfnamefont {X.}~\bibnamefont
  {Tong}},\ and\ \bibinfo {author} {\bibfnamefont {I.}~\bibnamefont
  {Zaliznyak}},\ }\bibfield  {title} {\bibinfo {title} {{Recent progress on
  HYSPEC, and its polarization analysis capabilities}},\ }\href
  {https://doi.org/10.1051/epjconf/20158303017} {\bibfield  {journal} {\bibinfo
   {journal} {EPJ Web of Conferences}\ }\textbf {\bibinfo {volume} {83}},\
  \bibinfo {pages} {03017} (\bibinfo {year} {2015})}\BibitemShut {NoStop}%
\end{thebibliography}%

\pagebreak
\pagebreak

\widetext
\begin{center}
\textbf{\large Supplemental Material for ``Easy-plane multi-$\mathbf{q}$ magnetic ground state of \ch{Na_3Co_2SbO_6}''}
\end{center}
\makeatletter
\renewcommand{\thefigure}{S\arabic{figure}}

\section{Single crystal growth}
Single crystals of \ch{Na_3Co_2SbO_6} were synthesized using a flux method \cite{LiPRX2022}. High-quality, twin-free single crystals were carefully selected through Raman spectroscopy. The selection took place at room temperature, employing parallel linear polarization for both incident and scattered photons on crystal surfaces parallel to the $ab$ plane \cite{LiPRX2022}. Two Raman scattering peaks between 200 and 220 cm$^{-1}$ displayed unique behavior when the polarization direction is parallel to the $\mathbf{b}$-axis. Furthermore, the twin-free nature of our crystal used for neutron diffraction on SENJU and SIKA was characterized by single-crystal $X$-ray diffraction measurements prior to the neutron diffraction experiments. The measurements were performed at room temperature with a custom-designed instrument equipped with a Xenocs Genix 3D Mo $K_\alpha$ (17.48 KeV) $X$-ray source. Three dimensional reciprocal map was achieved by sample rotation. All diffraction peaks could be indexed in a twin-free monoclinic setting \cite{LiPRX2022}. The crystal was later verified to be twin-free by the neutron diffraction data.

\bigbreak

\section{Unpolarized neutron diffraction experiments}
The experiment corresponding to Fig.~1 and Fig.~3 in the main text was conducted on the SENJU time-of-flight white beam neutron diffractometer at the MFL, J-PARC, Japan \cite{OhharaJAC2016}. A 6 mg twin-free crystal was mounted with reciprocal vectors $(H,-H,L)$ in the horizontal plane. In this configuration, the lower magnetic peaks in Fig. 3(b-e) and (g-i) were in the horizontal plane, while the upper peaks were not. The experiment utilized a 7 T vertical magnet, so that the fields were applied $30^\circ$ away from the $\mathbf{b}$ axis, as illustrated in Fig.~2 in the main text. All results were measured at 2 K following an initial zero-field cooling from 300 K.

The data presented in Fig.~3 were obtained after subtracting a background profile, which was determined for each pixel from its lowest intensity value across all different measurements under the same sample rotation angle. In measurements using a non-monochromated beam and a perfect single crystal for diffraction, the intensities of diffraction peaks produced by neutrons with different wavelengths $\lambda$ are affected by the Lorentz factor ($\propto d^2 \lambda^2$), where $d$ stands for $d$-spacing  \cite{MichelsClarkJAC2016}. The raw data has been processed through the software \textit{Stargazer}, which makes calibration on the white beam's intensity versus energy profile but does not correct for the Lorentz factor \cite{OhharaJAC2016}. Thus, the two magnetic peaks in Fig. 3(b-e) and (g-i) consistently displayed unequal intensities. To enable a quantitative comparison, the upper magnetic peak in Fig.~3(i) was produced by neutron with $\lambda \sim 1.24~\AA$, and the lower one with $\lambda \sim 3.5~\AA$. Consequently, the lower peak is expected to be approximately 8 times stronger than the upper one. The observed upper peak in Fig. 3(b-d) was further weakened because it fell partially outside the boundary of the detector bank. Later in our experiment, the sample rotation angle was adjusted slightly (by 2 degrees), which brought the peak back onto the detector bank. As a result, the upper magnetic peaks in Fig. 3(h-i) is stronger than in Fig. 3(b-d). In Fig.~3(i), the intensity ratio between the upper and lower peaks is consistent with the Lorentz factor within experimental error.

The above results have been corroborated by additional measurements on the same crystal using a triple-axis spectrometer, which allows us to closely track the field evolution of a single peak. The experiment was conducted on the SIKA cold neutron triple-axis spectrometer at the Australian Nuclear Science and Technology Organization (ANSTO) \cite{WuJOI2016}.  Measurements were performed using $k_\mathrm{i} = k_\mathrm{f} = 1.57$~\AA$^{-1}$ neutrons and a beryllium filter, maintaining the same sample and field geometry as in the SENJU experiment described above. All measurements were carried out at 2 K following zero-field cooling.

\begin{figure}[!ht]
\includegraphics[width=2.5in]{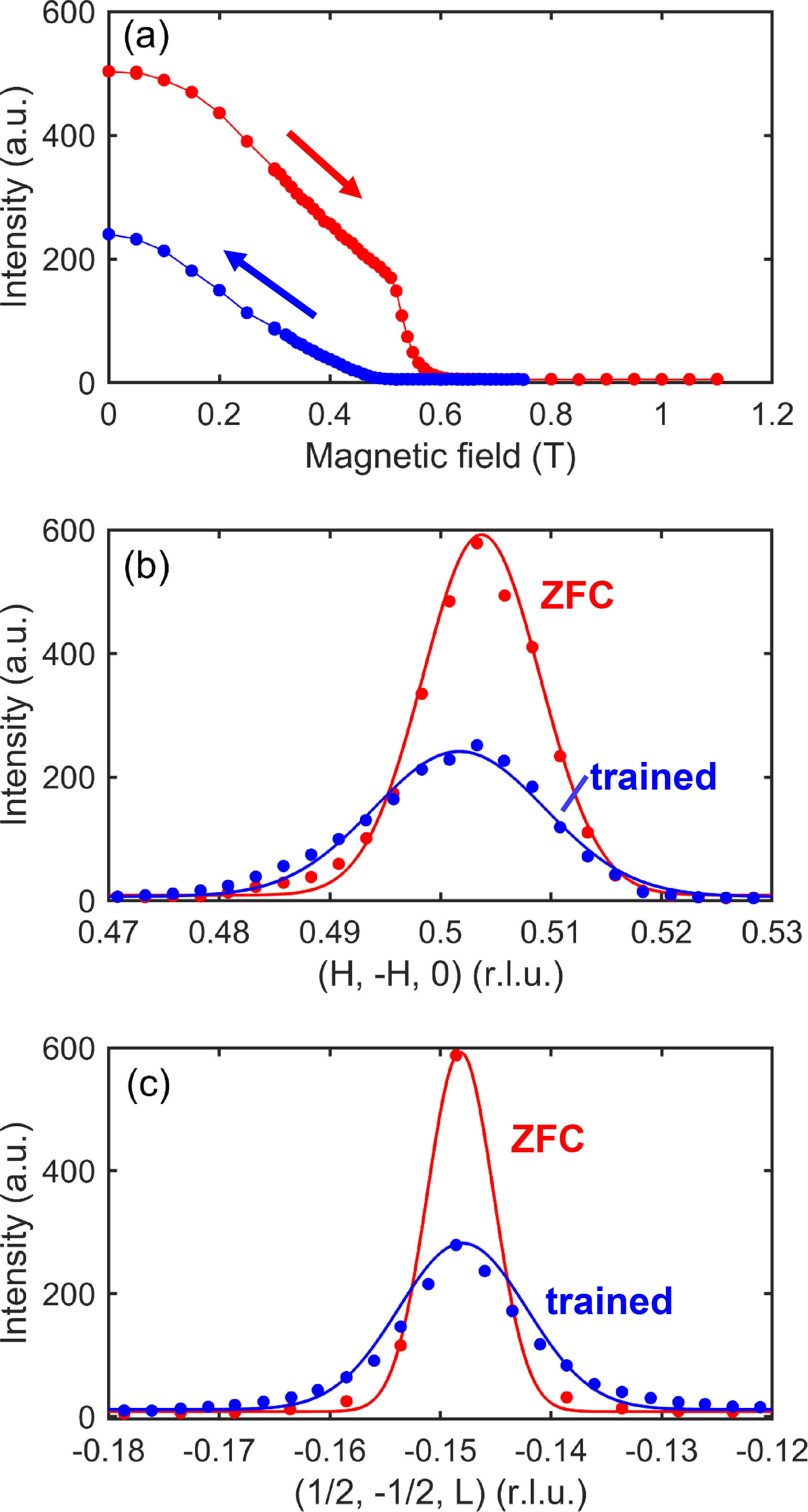}
\caption{(a) Field evolution of the $(1/2, -1/2 ,0)$ peak, measured at its center. (b-c) Momentum scans through the peak show noticeable broadening after training. Solid lines are Gaussian fits. After accounting for the broadening in both directions, the total weight of the peak is unchanged.}
\label{figS1}
\end{figure}

We selected the peak at $(1/2, -1/2, 0)$ for the measurement, which corresponds to the lower peak in Fig.~3b. In line with the findings from Fig. 3(c-d), the peak vanishes above 0.6 T [Fig.~\ref{figS1}(a)]. The crucial finding regarding field-training effects is the recovery of the peak upon removal of the field from above 1 T. Although the intensity measured at the peak center does not fully recover to the original value observed right after the zero-field cooling, it does not imply a genuine suppression of the associated zigzag domain or component by the field training. Instead, the intensity decrease results from a slight broadening of the peak, as shown in Fig.~\ref{figS1}(b-c). After accounting for this broadening, we conclude that the total weight of the peak remains unaffected by the training. A similar training-induced broadening has also been observed recently in \ch{Na_2Co_2TeO_6} \cite{YaoArxiv2022}. In Fig.~\ref{figS2}, we demonstrate that the magnetization response of another twin-free crystal is independent of a zero-field-cooling or field-cooling history. All these results conclusively show an absence of field-training effects and identify the zero-field ground state as having double-$\mathbf{q}$ characteristics.

\bigbreak

\section{Magnetization before and after field training}
\begin{figure}[!ht]
\includegraphics[width=3in]{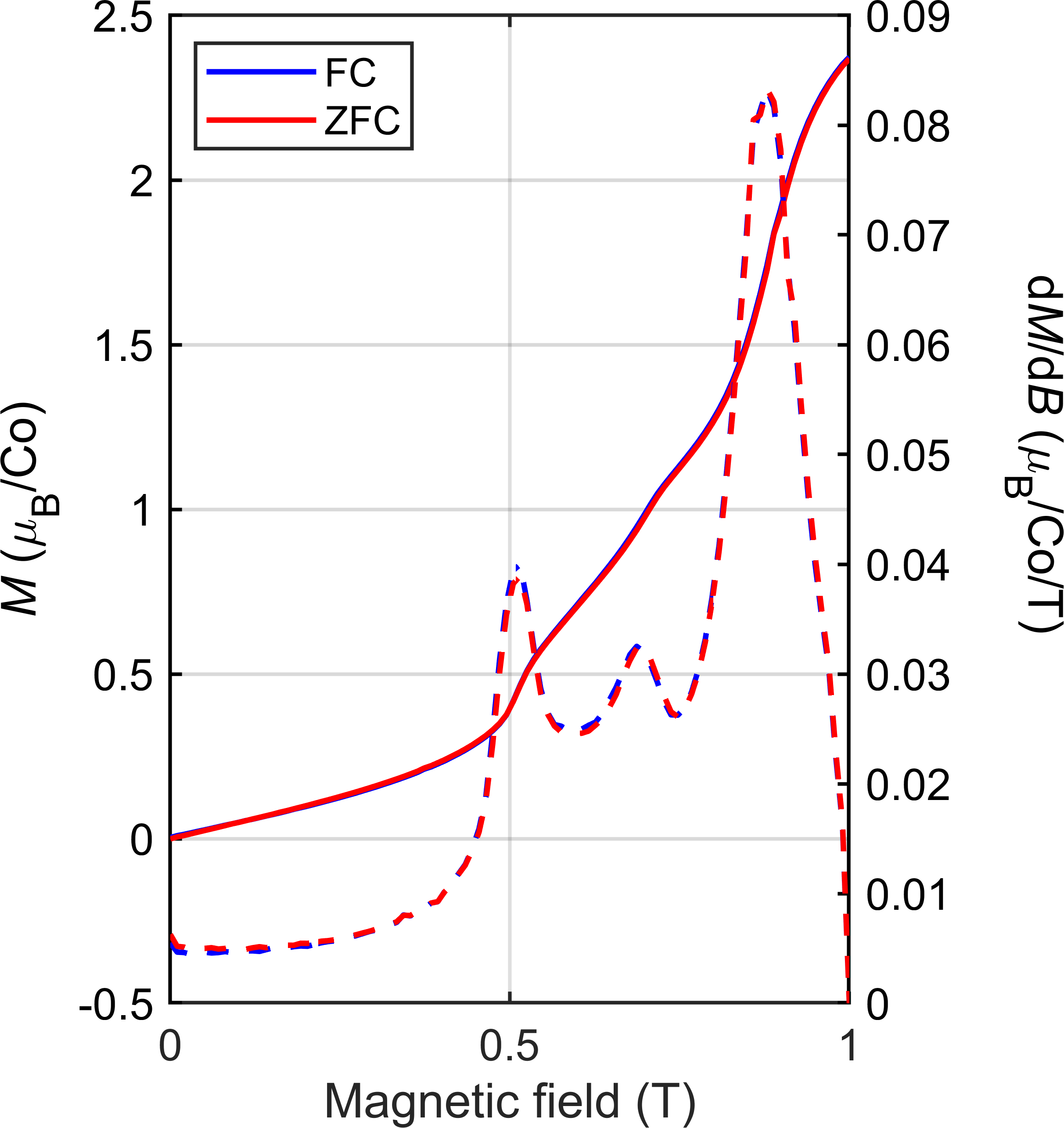}
\caption{Magnetization (solid line) and field derivative (dashed line) measured on a twin-free crystal at 2 K with the field applied 30 degrees from the $b$-axis. Zero-field and field (7000 Oe) cooling from 300 K produce the same result.}
\label{figS2}
\end{figure}

DC magnetometry was performed with a Quantum Design MPMS on a twin-free single crystal of mass about 2.46 mg, using a quartz rod as sample holder. In order to determine whether in-plane fields applied at 30 degrees away from the $\mathbf{b}$ axis can produce a noticeable difference in the magnetic (zigzag) domain population, we compared measurements performed after both zero-field cooling (ZFC) and field cooling (FC). The ZFC result was measured after a zero-field cooling from 300 K to 2 K, whereas the FC result was measured after cooling from 300 K to 2 K in a 7000 Oe field (followed by a linear field removal). The magnetization measurements were performed in field increments of 100 Oe, and we used continuous sweeping mode to avoid field oscillations. The direction of the applied fields were the same as for our twin-free-crystal neutron diffraction experiments described above. As shown in Fig.~\ref{figS2}, no difference was observed between the ZFC and FC data.

\bigbreak

\section{Spin-polarized neutron diffraction}
Our spin-polarized neutron diffraction measurement was performed on the HYSPEC time-of-flight spectrometer at the SNS, Oak Ridge National Laboratory \cite{BarryEPJ2015}. The sample was cooled to 0.3 K using a helium-3 insert placed inside a compensated and asymmetric 5 Tesla vertical field magnet. The twinned sample array had a total mass of 0.5 grams and a full mosaic spread of about $2.3^\circ$. Using an incident neutron energy of $E_\mathrm{i} = 5.3$ meV, the narrow vertical opening angle (about $\pm7^\circ$) of the set up essentially allowed for detection of scattering signals only inside the horizontal scattering plane. The diffraction data were collected by rotating the sample about the vertical axis over a $96^\circ$ range in $1^\circ$ steps. This process yielded a two-dimensional diffraction dataset, which we further symmetrized using mirror operations (as displayed in Fig.~\ref{figS4}). Data were corrected by a polarization flipping ratio of 11.3 \cite{ZaliznyakIOP2017}, which optimally removed the nuclear peaks from the spin-flip channel.

\begin{figure*}[!ht]
\includegraphics[width=5.6in]{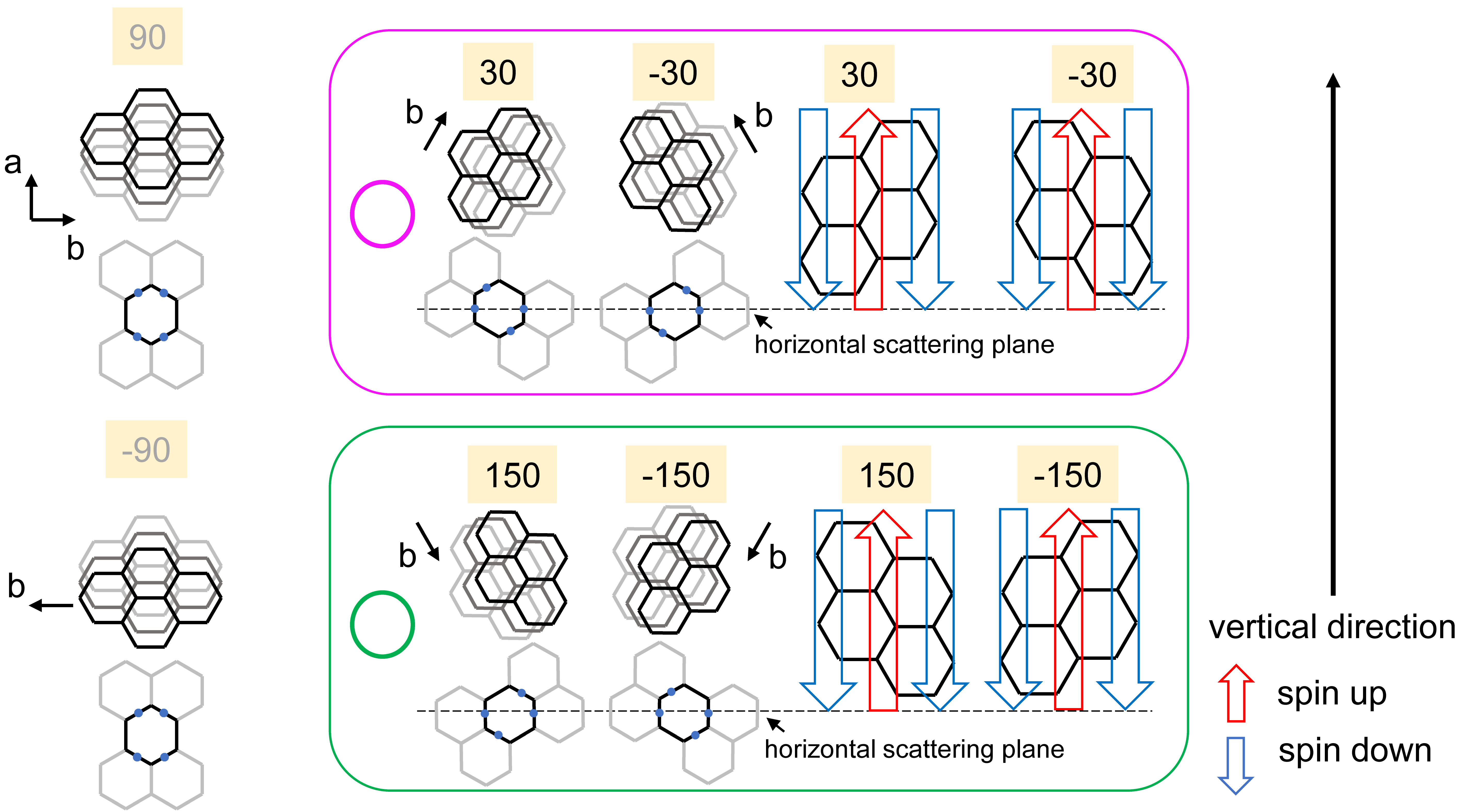}
\caption{A schematic of twin domains and our scattering geometry. The crystallographic domains are identified by the angle spanned by the $\mathbf{b}$-axis and vertical field direction. The associated 2D Brillouin zones are pair-wise displayed under the real-space picture, with the $M$-point magnetic Bragg peaks indicated on the zone boundary. The four domains within the green and magenta blocks contribute to the diffraction signals inside the horizontal scattering plane. Since only non-spin-flip signals are observed (see Fig.~4 of main text), the staggered spins must be in the vertical direction (as indicated by empty arrows).}
\label{figS3}
\end{figure*}

\begin{figure}[!ht]
\includegraphics[width=3in]{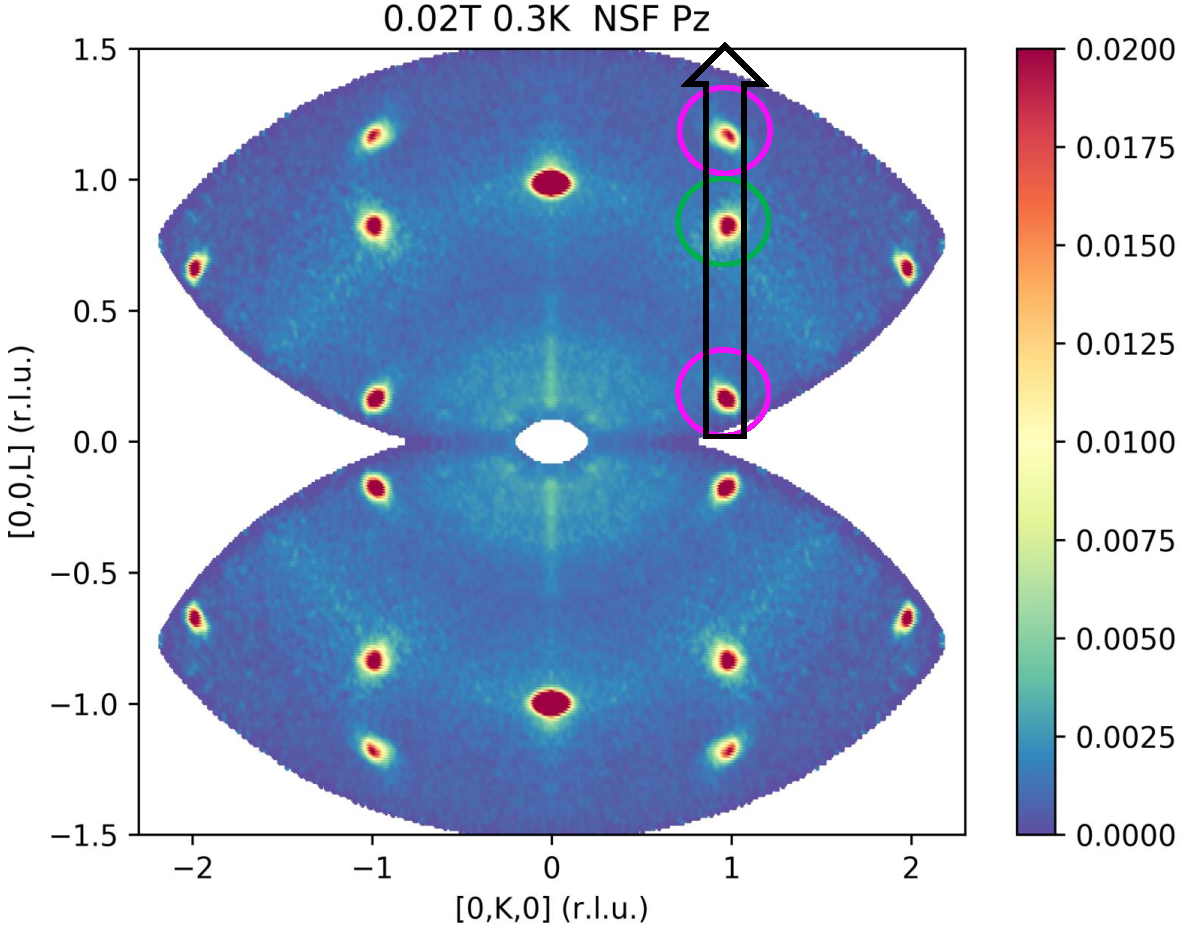}
\caption{The non-spin-flip neutron diffraction results inside horizontal scattering plane. The line-cut in Fig.~4 is indicated by the arrow, which goes through three different magnetic Bragg peaks. The domains contributing to these peaks are indicated by circles color-coded after Fig.~\ref{figS3}.}
\label{figS4}
\end{figure}

The experiment used a fully-twinned array of crystals. As depicted in Fig.~\ref{figS3}, we identify six distinct domain types, each corresponding to a different monoclinic stacking direction with respect to the horizontal scattering plane. Four of the domains (``30'', ``-30'', ``150'', and ``-150'') have magnetic diffraction peaks in the scattering plane, whereas the remaining two (``90'' and ``-90'') have not. The data in Fig.~4 of main text are obtained from a line cut through the data in Fig.~\ref{figS4}. The magnetic peaks are indicated with color-coded circles according to Fig.~\ref{figS3} to indicate their domain origin. Whenever a zigzag component produces a magnetic reflection in the horizontal scattering plane, we can analyze the spin orientation of its staggered moments, as explained in the main text. The essence of the result is that all of the detected zigzag components feature (staggered) magnetization in the vertical direction, as illustrated by the empty arrows in Fig.~\ref{figS3}. Associating the observed zigzag components with the underlying crystal lattice and superposing them together, we obtain the double-$\mathbf{q}$ magnetic structure illustrated in Fig.~2 of the main text. Our spin-polarized diffraction results are inconsistent with a previously proposed spin configuration for NCSO with all spins parallel to the $\mathbf{b}$ axis \cite{YanPRM2019}, because that would produce non-zero spin-flip scattering signals which are not observed in our data.

\pagebreak	

\section{Supplementary Figures}

\end{document}